\documentclass[prd,amsmath,amssymb,nofootinbib,balancelastpage,longbibliography,twocolumn]{revtex4-1}
\usepackage{multirow}
\usepackage{epsfig}
\usepackage{amsmath}
\usepackage{bm}
\usepackage{times}
\usepackage{graphicx}
\usepackage{color}
\usepackage{slashed}
\usepackage{graphicx}
\usepackage{amsmath}
\usepackage{tikz}
\usetikzlibrary{positioning,shapes}
\usepackage{relsize} 
\usepackage[latin1]{inputenc}
\usepackage{hyperref}
\usepackage{soul} 

\def\bea{\begin{eqnarray}}
\def\eea{\end{eqnarray}}
\def\bean{\begin{equation*}}
\def\eean{\end{equation*}}

\def\bea{\begin{eqnarray}}
\def\eea{\end{eqnarray}}
\def\bean{\begin{equation*}}
\def\eean{\end{equation*}}

\begin{document}

\title{Baryon Number and Lepton Universality Violation in Leptoquark and Diquark Models}

\author{Nima~Assad}
\author{Bartosz~Fornal}
\affiliation{Department of Physics, University of California, San Diego, 9500 Gilman Drive, La Jolla, CA 92093, USA}
\author{Benjam\'{i}n~Grinstein}
\affiliation{Department of Physics, University of California, San Diego, 9500 Gilman Drive, La Jolla, CA 92093, USA}
\date{\today}

\begin{abstract}
We perform a systematic study of models involving leptoquarks and diquarks with masses well below the grand unification scale and demonstrate that a large class of them is excluded due to rapid proton decay. After singling out the  few phenomenologically viable color triplet and sextet scenarios, we show that there exist only two leptoquark models which do not suffer from tree-level proton decay and which have the potential for explaining the recently discovered anomalies in $B$ meson decays. Both of those models, however, contain dimension five operators contributing to proton decay and require a new symmetry forbidding them to emerge at a higher scale. This has a particularly nice realization for the model with the vector leptoquark $(3,1)_{2/3}$, which points to a specific extension of the Standard Model, namely the Pati-Salam unification model, where this leptoquark naturally arises as the new gauge boson. We explore this possibility in light of recent $B$ physics measurements. Finally, we analyze also a vector diquark model, discussing its LHC phenomenology and showing that it has nontrivial predictions for neutron-antineutron oscillation experiments.
\vspace{9mm}
\end{abstract}

\maketitle

\section{Introduction}  
Protons have never been observed to decay. Minimal  grand unified theories (GUTs) \cite{Georgi:1974sy,Fritzsch:1974nn} predict proton decay at a rate which should have already been measured. The only four-dimensional GUTs constructed so far  based on a single unifying gauge group with a stable proton require either imposing specific gauge conditions \cite{Karananas:2017mxm} or introducing new particle representations \cite{Fornal:2017xcj}. A detailed  review of the subject can be found in \cite{Nath:2006ut}. Lack of experimental evidence for proton decay \cite{Miura:2016krn} imposes severe constraints on the form of new physics, especially on theories involving new bosons with masses well below the GUT scale.  For phenomenologically viable models of physics beyond the Standard Model (SM) the new particle content cannot trigger fast proton decay, which seems like an obvious requirement, but is often ignored in the model building literature. 

Simplified models with additional  scalar leptoquarks and diquarks not triggering tree-level proton decay were discussed in detail in \cite{Arnold:2012sd}, where a complete list of color singlet, triplet and sextet scalars coupled to fermion bilinears was presented. An interesting point of that analysis is that there exists only one scalar leptoquark, namely $(3,2)_{\frac76}$ (a color triplet electroweak doublet with hypercharge $7/6$) that does not cause tree-level proton decay. In this model dimension five operators that mediate proton decay can be forbidden by imposing an additional symmetry \cite{Arnold:2013cva}.

In this paper we collect the results of \cite{Arnold:2012sd} and extend the analysis to vector particles. This scenario might be regarded as more appealing than the scalar case, since the new fields do not contribute to the hierarchy problem. We do not assume any global symmetries, but we do comment on how imposing a larger symmetry can remove  proton decay that is introduced through nonrenormalizable operators, as in the scalar case.

Since many models for the recently discovered $B$ meson decay anomalies \cite{Aaij:2014ora,Aaij:2017vbb} rely on the existence of new scalar or vector leptoquarks, it is interesting to investigate which of the new particle explanations proposed in the literature do not trigger rapid proton decay. Surprisingly, the requirement of no proton decay at tree level singles out only a few models, two of which  involve the vector leptoquarks $(3,1)_{\frac23}$ and  $(3,3)_{\frac23}$, respectively. Remarkably, these very same representations have been singled out as giving better fits to $B$ meson decay anomalies data~\cite{Alonso:2015sja}. An interesting question we consider is whether there exists  a UV complete extension of the SM containing such leptoquarks in its particle spectrum.

Finally, although the phenomenology of leptoquarks has been analyzed in great detail, there still remains a gap in the discussion of diquarks. In particular,  neutron-antineutron ($n - \bar{n}$) oscillations have not been considered in the context of vector diquark models.  We fill  this gap by deriving an estimate for the $n - \bar{n}$ oscillation rate in a simple vector diquark model and discuss its implications for present and future experiments.

The paper is organized as follows. In Sec.~\ref{repr} we  study the order at which proton decay first appears in models including new color triplet and sextet representations and briefly comment on their experimental status. In Sec.~\ref{leptvec} we focus on the unique vector leptoquark model which does not suffer from tree-level proton decay and has an appealing UV completion. In particular, we study its implications for $B$ meson decays. In Sec.~\ref{other} we analyze a model with a single vector color sextet, discussing its LHC phenomenology and implications for $n - \bar{n}$ oscillations.  Section~\ref{conclusions} contains conclusions.

{\renewcommand{\arraystretch}{1.8}\begin{table}[b!]
\begin{center}\vspace{0mm}
    \begin{tabular}{| c | c |}
    \hline
       Field   & \ \ ${\rm SU}(3)_c\times {\rm SU}(2)_L\times {\rm U}(1)_Y$ reps.  \ \
       \\ [1pt] \hline\hline
                           Scalar leptoquark & \ \ $\left(3,2\right)_{\frac76}'$  \ \  \\ [2pt] \hline                               
           \ \ {Scalar diquark}  \ \  &  \  $\left(3,1\right)_{\frac23}$, $\left(6,1\right)_{-\frac23}$, $\left(6,1\right)_\frac13$, $\left(6,1\right)_{\frac43}$, $\left(6,3\right)_\frac13$  \ \\ [2pt] \hline 
                \  \    Vector leptoquark  \  \ & \ \ $\left(3, 1\right)_\frac23'$, $\left(3,1\right)_\frac53$, $\left(3, 3\right)_\frac23'$  \ \  \\ [2pt] \hline                               
           \ \ Vector diquark  \ \  & \ \ $\left(6,2\right)_{-\frac16}$, $\left({6},2\right)_{\frac56}$  \ \  \\ [2pt] \hline                                    
    \end{tabular}
\end{center}
\vspace{-2mm}
\caption{\small{The only leptoquark and diquark models with a triplet or sextet color structure that do not suffer from tree-level proton decay. The primes indicate the existence of dim 5 proton decay channels.}}
\label{table2}
\end{table}}

\section{Viable leptoquark and diquark models}
\label{repr}

For clarity, we first summarize the combined results of \cite{Arnold:2012sd}  and this work in Table~\ref{table2}, which shows the only color triplet and color sextet models that do not exhibit tree-level proton decay. The  scalar case was investigated  in \cite{Arnold:2012sd}, whereas in this paper we concentrate on vector particles. As explained below, the representations denoted by primes  exhibit proton decay through dimension five operators (see also \cite{Arnold:2013cva}).  We note that although the renormalizable proton decay channels involving leptoquarks are well-known in the literature, to our knowledge the nonrenormalizable channels have not been considered anywhere  apart from the scalar case in \cite{Arnold:2013cva}.

\subsection{Proton decay in vector models}
\begin{figure}[t!]
\includegraphics[width=0.6\linewidth]{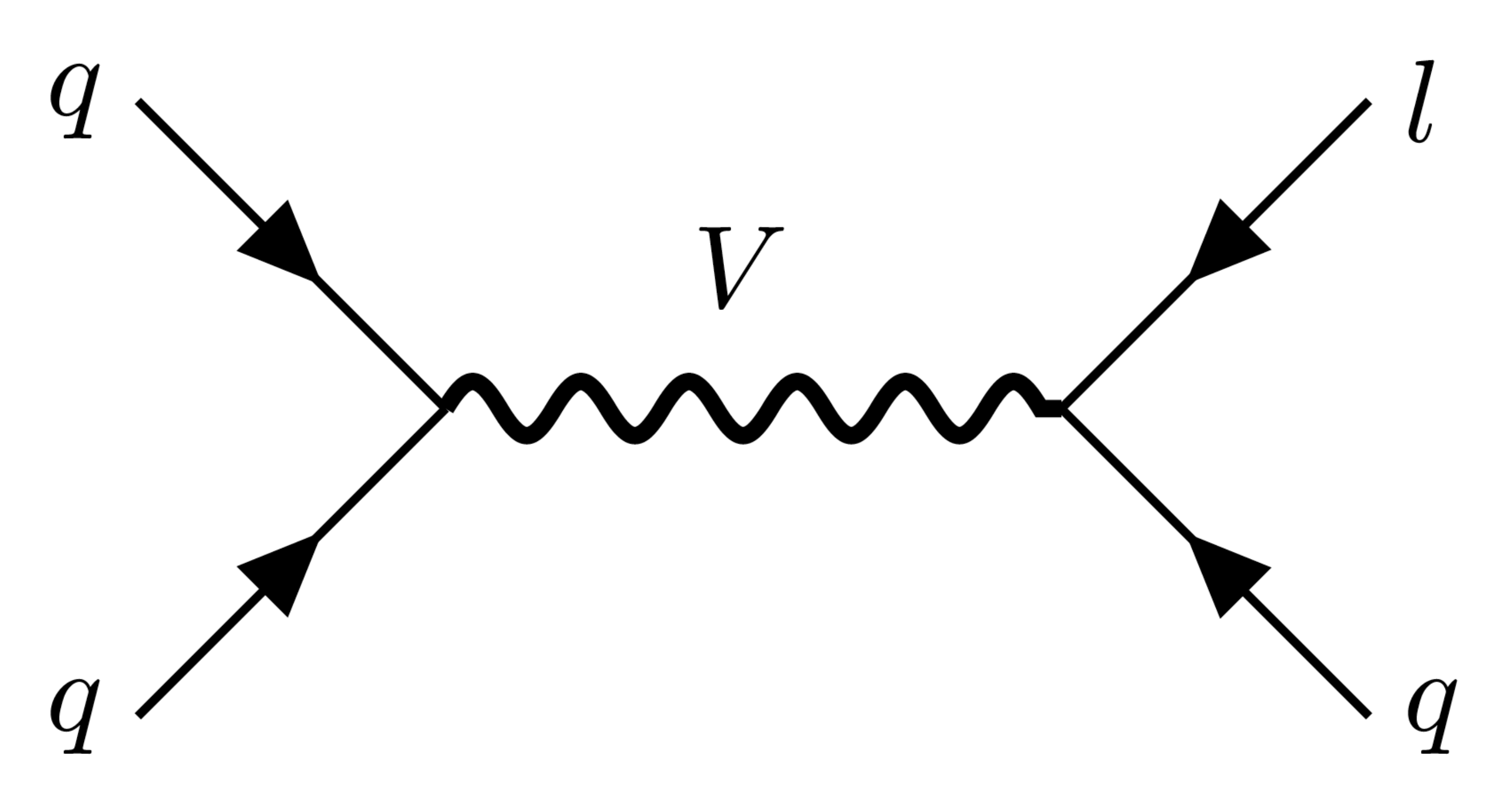} \vspace{-1mm}
\caption{\small{Tree-level vector exchange triggering proton decay for \\ \centerline{ $ V =(3,2)_{\frac16}$ and  $V=(3,2)_{-\frac56}$.}}}\vspace{3mm}
\label{fig:1}
\end{figure}

\begin{figure}[t!]
\includegraphics[width=0.64\linewidth]{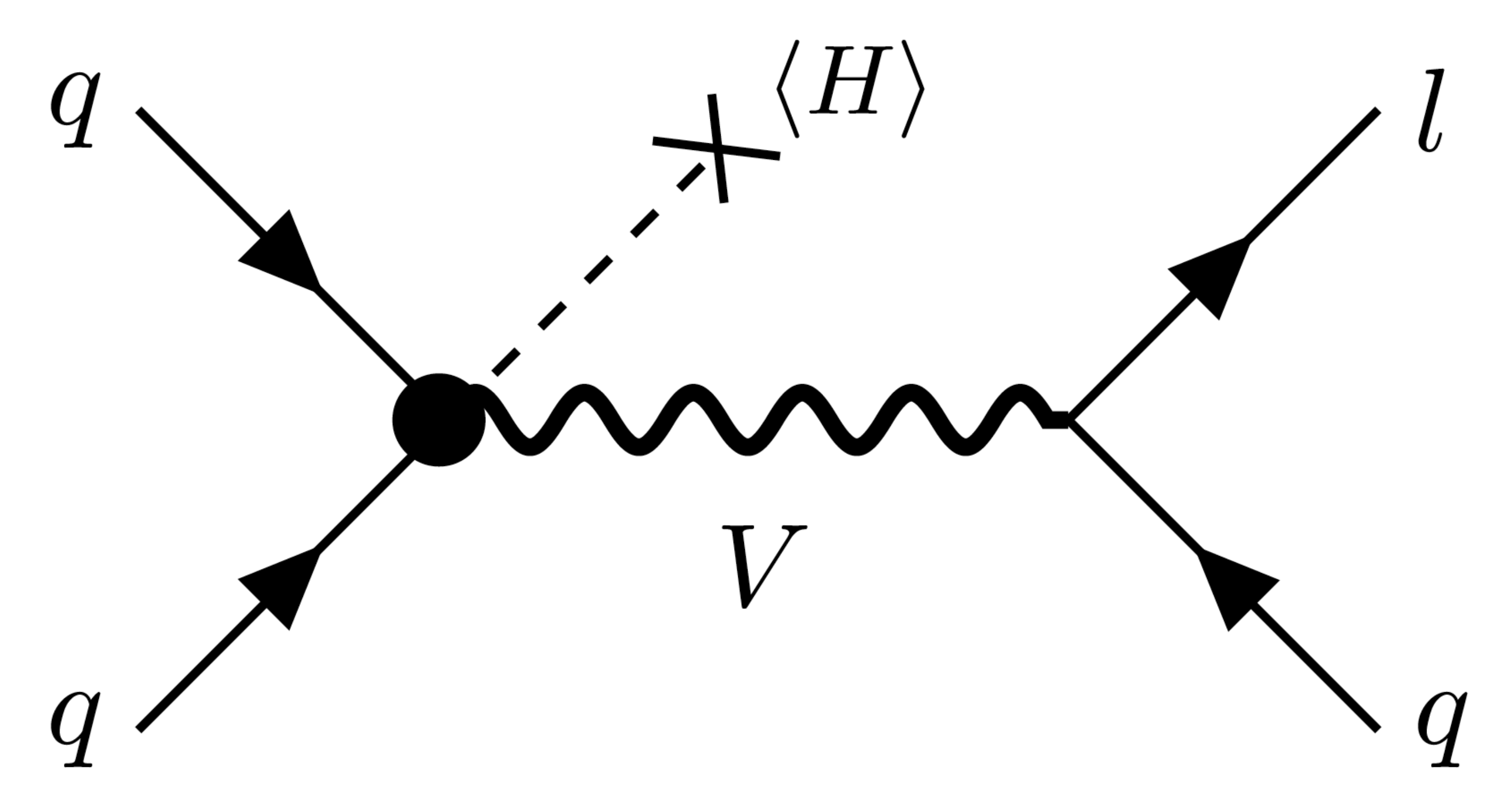} \vspace{-1mm}
\caption{
\small{Proton decay through a dimension five interaction for \\ \centerline{ $ V =(3,1)_{\frac23}$ and  $V=(3,3)_{\frac23}$.}}}
\label{fig:2}
\end{figure}

We first enumerate all possible dimension four interactions of the new vector color triplets and sextets with fermion bilinears respecting  gauge and Lorentz invariance. A complete set of those  operators is listed in Table~\ref{table1}~\cite{Dorsner:2016wpm}. For the vector case there are two sources of proton decay. The first one comes from tree-level diagrams involving a vector color triplet exchange, as shown in Fig.~\ref{fig:1}. This excludes the representations $(3, 2)_\frac16$ and $(3, 2)_{-\frac56}$, since  they would require unnaturally small couplings to SM fermions or very large masses to remain consistent  with proton decay limits. 
The second source  comes from dimension five operators involving the vector leptoquark representations $(3, 1)_\frac23$ and $(3, 3)_\frac23$:
\begin{equation}\label{dim5op}
\frac{1}{\Lambda}\, (\overline{Q}^c_L H^\dagger)\gamma^\mu d_R V_\mu \,  ,  \ \ \ \ \frac{1}{\Lambda}\, (\overline{Q}^c_L \tau^A H^\dagger)\gamma^\mu d_R V^A_\mu  \ ,
\end{equation}
respectively. Those operators can be constructed if no additional global symmetry forbidding them is imposed and allow for the proton decay channel shown in Fig.~\ref{fig:2}, resulting in a lepton (rather than an antilepton) in the final state. The corresponding proton lifetime  estimate is:
\begin{equation}   \label{pl}
\tau_p \approx \left(2.5\times 10^{32} \ {\rm years}\right) \left(\frac{M}{10^4 \ {\rm TeV}}\right)^4 \left(\frac{\Lambda}{M_{\rm Pl}}\right)^2, \ 
\end{equation}
where the leptoquark tree-level coupling and the coefficient of the dimension five operator were both set to unity. The numerical factor in front of Eq.~(\ref{pl}) is the current  limit on the proton lifetime from the search for  $p\to K^+\pi^+e^-$ \cite{Olive:2016xmw}. Even in the most optimistic scenario of the largest suppression of proton decay, i.e., when the new physics behind the dimension five operator  does not appear below the Planck scale, those operators are still problematic for $M \lesssim 10^4 \ {\rm TeV}$, which well includes the region of interest for the $B$ meson decay anomalies.

The dimension five operators can be removed by embedding the vector leptoquarks into UV complete models. As  argued in \cite{Arnold:2013cva} for the scalar case, it is sufficient to impose a discrete subgroup $\mathcal{Z}_3$ of a global ${\rm U}(1)_{B-L}$ to forbid the problematic dimension five operators. They are also naturally absent in models with gauged ${\rm U}(1)_{B-L}$.\footnote{We note that the dimension five operators (\ref{dim5op})  provide a baryon number violating channel which may be used to generate a cosmological baryon number asymmetry.}

Ultimately, as shown in Table~\ref{table2}, there are only five color triplet or sextet vector representations that are free from tree-level proton decay, two of which  produce dimension five proton decay operators. 
In the scalar case, as shown in  \cite{Arnold:2012sd}, there are six possible representations with only one suffering from dimension five proton decay. 

{\renewcommand{\arraystretch}{1.4}\begin{table}[t!]
\begin{center}
    \begin{tabular}{| c || c | c | c | c|}
    \hline
       \ \ Operator \ \  & \raisebox{0ex}[0pt]{$ \ \ {\rm SU}(3)_c \ \ $} & \raisebox{0ex}[0pt]{$ \ \  {\rm SU}(2)_L \ \ $} & \raisebox{0ex}[0pt]{$ \ \ \ {\rm U}(1)_Y \ \ \ $}  & \ \ \ $p$ decay \ \ \ 
       \\ \hline\hline   
      \ \ \raisebox{-2.5mm}{$\overline{Q}^c_L \gamma^\mu u_R V_\mu   $} \ \  & $3 $ & $2$ & $-\,5/6 \ \ $ & tree-level \\ [-5pt] \cline{2-5}
          & $\bar{6}$ & $2$ & $-\,5/6 \ \ $ & --  \\ \hline
       \ \ \raisebox{-2.5mm}{$\overline{Q}^c_L \gamma^\mu d_R V_\mu   $} \ \  & $3 $ & $2$ & $ 1/6  $ & tree-level \\ [-5pt] \cline{2-5}
          & $\bar{6}$ & $2$ & $1/6 $ & -- \\ \hline
                \ \  $\overline{Q}_L \gamma^\mu L_L V_\mu $ \ \  & $3 $ & $1 , 3$ & $2/3 $ & dim 5\\ \hline
                        \ \  $\overline{Q}^c_L \gamma^\mu {e}_R V^*_\mu $ \ \  & ${3}$ & $2$ & $-\,5/6 \ \  $ & tree-level \\ \hline
                        \ \  $\overline{L}^c_L\gamma^\mu {u}_R V^*_\mu $ \ \  & ${3}$ & $2$ & $1/6  $ & tree-level\\ \hline       
                        \ \  $\overline{L}^c_L\gamma^\mu {d}_R V^*_\mu $ \ \  & ${3}$ & $2$ & $-5/6 \ \  $ & tree-level \\ \hline    
                \ \  $\overline{u}_R\gamma^\mu {e}_R V_\mu $ \ \  & ${3}$ & $1$ & $5/3  $ & dim 7 \\ \hline   
                 \ \  $\overline{d}_R\gamma^\mu {e}_R V_\mu $ \ \  & ${3}$ & $1$ & $2/3  $ & dim 5\\ \hline      
    \end{tabular}
\end{center}\vspace{-2mm}
\caption{\small{Possible vector color triplet and sextet representations.}}\vspace{1mm}
\label{table1}
\end{table}}

\subsection{Leptoquark phenomenology} \label{qqq}
The phenomenology of scalar and vector leptoquarks  has been extensively discussed in the literature \cite{Buchmuller:1986zs,Davidson:1993qk,Hewett:1997ce} and we do not attempt to provide a complete list of all relevant papers here. For an excellent review and many references see \cite{Dorsner:2016wpm}, which is focused primarily on light leptoquarks. 

Low-scale leptoquarks have recently become a very active area of research due to their potential for explaining the experimental hints of new physics in $B$ meson decays, in particular $B^+ \rightarrow K^+ \ell^+ \ell^-$ and $B^0 \rightarrow K^{*0} \ell^+ \ell^-$, for which a deficit in  the ratios $R_{K^{(*)}}=\text{Br}(B\!\to\! K^{(*)}\mu^+\mu^-)/\text{Br}(B\!\to\! K^{(*)}e^+e^-)$ with respect to the  SM expectations has been reported \cite{Aaij:2014ora,Aaij:2017vbb}. A detailed analysis of the anomalies can be found in \cite{Geng:2017svp,Ciuchini:2017mik,Hiller:2017bzc,DAmico:2017mtc,Altmannshofer:2017yso,Capdevila:2017bsm}.  Several leptoquark models have been proposed to alleviate this tension and are favored by a global fit to $R_{K^{(*)}}$,~$R_{D^{(*)}}$ and other flavor observables. Surprisingly, not all of those models remain free from tree-level proton decay.

The leptoquark models providing the best fit to data with just a single new representation are: scalar $(3,2)_{\frac16}$  \cite{Becirevic:2016yqi}, vector $(3,1)_{\frac23}$ \cite{Alonso:2015sja,Buttazzo:2017ixm} and vector $(3,3)_{\frac23}$ \cite{Fajfer:2015ycq}. Among those, only the models with the vector leptoquarks $(3,1)_{\frac23}$ and $(3,3)_{\frac23}$ are naturally free from any tree-level proton decay, since for the scalar leptoquark $(3,2)_{\frac16}$ there exists a dangerous quartic coupling involving three leptoquarks and the SM Higgs \cite{Arnold:2012sd} which triggers tree-level proton decay. 

Interestingly, as indicated in Table~\ref{table1}, both vector models  $(3,1)_{\frac23}$ and $(3,3)_{\frac23}$ suffer from dimension five proton decay and require imposing an additional symmetry to eliminate it. An elegant way to do it would be to extend the SM symmetry by a gauged ${\rm U}(1)_{B-L}$. Actually, such an extended symmetry would eliminate also the tree-level proton decay in the model with the scalar leptoquark $(3,2)_{\frac16}$.
 However, as we will see  in Sec.~\ref{leptvec},  only in the case of the:
 \begin{itemize}
\item[$\blacktriangleright$] \ \  vector leptoquark $(3,1)_{\frac23}$
\end{itemize}
there exists a very appealing SM extension which intrinsically contains such a state in its spectrum, simultaneously forbidding dimension five proton decay.

\subsection{Diquark phenomenology}\label{dq}
The literature on the phenomenology of diquarks is much more limited. It focuses on scalar diquarks \cite{Hewett:1988xc} and predominantly looks at three aspects:  LHC discovery reach for scalar diquarks \cite{Tanaka:1991nr,Atag:1998xq,Cakir:2005iw,Chen:2008hh,Han:2009ya,Gogoladze:2010xd,Berger:2010fy,Baldes:2011mh,Richardson:2011df,Karabacak:2012rn,Kohda:2012sr,Das:2015lna,Chivukula:2015zma,Zhan:2013sza}, $n - \bar{n}$ oscillations mediated by scalar diquarks \cite{Mohapatra:1980qe,Babu:2008rq,Ajaib:2009fq,Gu:2011ff,Baldes:2011mh,Babu:2012vc,Arnold:2012sd,Babu:2013yca} and baryogenesis \cite{Babu:2008rq,Arnold:2012sd,Baldes:2011mh,Gu:2011ff,Babu:2012vc,Babu:2013yca}. Studies of vector diquarks investigate only their LHC phenomenology \cite{Arik:2001bc,Sahin:2009dca,Richardson:2011df,Zhang:2010kr,Grinstein:2011yv,Grinstein:2011dz},  concentrating on their interactions with quarks.

In Sec.~\ref{other} we close the gap  in diquark phenomenology by discussing the implications of a vector diquark model for $n - \bar{n}$ oscillation experiments.

\section{Vector leptoquark model}\label{leptvec}

As emphasized in Sec.~\ref{qqq}, the SM extended by just the vector leptoquark  $(3,1)_{\frac23}$  is a unique model, which,  apart from being free from tree-level proton decay, has a very simple and attractive UV completion, automatically forbidding dimension five proton decay operators. 

\subsection{Pati-Salam unification}

\vspace{-1mm}
A priori, any of the leptoquarks can originate from an extra GUT irrep, either scalar or vector. In particular,  the vector leptoquark  $(3,1)_{\frac23}$ could be a component of the vector $40$ irrep of ${\rm SU}(5)$. Nevertheless, this generic explanation does not seem to be strongly motivated or predictive. Another interpretation of the vector $(3,1)_{\frac23}$ state arises in composite models \cite{Gripaios:2009dq,Gripaios:2014tna,Barbieri:2016las}.  

The third, perhaps the most desirable option, is that the vector leptoquark $(3,1)_{\frac23}$ is the gauge boson of a unified theory. Indeed, this scenario is realized  if one considers partial unification based on the Pati-Salam gauge group:
\begin{equation}
{\rm SU}(4) \times {\rm SU}(2)_L \times {\rm SU}(2)_R
\end{equation}
at higher energies  \cite{Pati:1974yy}.\footnote{Similar unification models have recently been constructed based on the gauge group ${\rm SU}(4) \times {\rm SU}(2)_L \times {\rm U}(1)$ \cite{Perez:2013osa,Fornal:2015boa}, in which the new gauge bosons also coincide with the vector leptoquark $(3,1)_{2/3}$.\vspace{1mm}} In this case  the vector leptoquark $(3,1)_{\frac23}$ emerges naturally as the new gauge boson of the broken symmetry, and is completely independent of the symmetry breaking pattern. 
It is also interesting that the Pati-Salam partial unification model can be fully unified into an ${\rm SO}(10)$ GUT.

The fermion  irreps of the Pati-Salam model, along with their decomposition into SM fields, are \vspace{0mm}
\begin{equation} 
\label{eq:PatiSalamIrreps}
\begin{aligned}
(4,2,1) &= (3,2)_{\frac16} \oplus (1,2)_{-\frac12}\\
(\bar4,1,2) &= (\bar3,1)_{\frac13} \oplus  (\bar3,1)_{-\frac23} \oplus (1,1)_1 \oplus (1,1)_0\,.
\end{aligned}
\end{equation} 
Interestingly, the theory is free from tree-level proton decay via gauge interactions. The explanation for this is straightforward.
Since ${\rm SU}(4) \supset {\rm SU}(3)_c \times {\rm U}(1)_{B-L}$, this implies that $B\!-\!L$ is conserved. However, after the Pati-Salam group breaks down to the SM, the interactions of the leptoquark $(3,1)_{\frac23}$ with quarks and leptons have an accidental $B\!+\!L$ global symmetry. Those two symmetries combined result in both baryon and lepton number being conserved in gauge interactions, thus no proton decay can occur via a tree-level exchange of $(3,1)_{\frac23}$. 

In addition, there are  no gauge-invariant dimension five proton decay operators in the Pati-Salam model involving  the vector leptoquark $(3,1)_{\frac23}$. This was actually expected from the fact that ${\rm SU}(4) \supset {\rm SU}(3)_c \times {\rm U}(1)_{B-L}$ and, as discussed in Sec.~\ref{qqq}, a ${\rm U}(1)_{B-L}$ symmetry is sufficient to forbid such operators.

\subsection{Flavor structure}
The quark and lepton mass eigenstates are related to the gauge eigenstates through $n_f\times n_f$ unitary matrices, with $n_f=3$ the number of families of quarks and leptons. Expressing the interactions  that couple the $(3,1)_{\frac23}$ vector leptoquark to the quark and the lepton in each irrep of Eqs.~\eqref{eq:PatiSalamIrreps} in terms of mass eigenstates,  one must include unitary matrices, similar to the Cabibbo-Kobayashi-Maskawa (CKM) matrix for the quarks in the SM, that measure the misalignment of the lepton and quark mass eigenstates:
\begin{align}
\label{eq:newINT}
\mathcal{L} \ \supset \  \frac{g_4}{\sqrt2}\, V_\mu \,&\Big[L^u_{ij}\,(\bar u^i\gamma^\mu P_L\nu^j)+L^d_{ij}\,(\bar d^i \gamma^\mu P_L e^j)
\nonumber\\
&+\ R_{ij}\,(\bar d^i\gamma^\mu P_R \, e^j)\Big] + {\rm h.c.}\,.
\end{align}
The ${\rm SU}(4)$ gauge coupling constant,  $g_4$, is not an independent paramter but fixed by the QCD coupling constant at the scale $M$ of the masses of the vector bosons of ${\rm SU}(4)$; to leading order $g_4(M)=\!\sqrt{4\pi \alpha_s(M)}\approx0.94$ at $M=16$~TeV, where, as will become clear later, this is the maximum value of $M$ consistent with the $R_{K^{(*)}}$ anomaly  in this model. The unitary matrices $L^u$  and $L^d$, the  CKM matrix $V$   and the  Pontecorvo-Maki-Nakagawa-Sakata (PMNS) matrix $U$  satisfy $L^u= VL^dU$. 

\subsection{B meson decays}
In the SM, flavor-changing neutral currents with $\Delta B=-\Delta S=1 $  are described by the effective Lagrangian~\cite{Grinstein:1988me,Buchalla:1995vs}:
\begin{equation}
\label{eq:Leffnc}
  {\cal{L}}= -
  \frac{4 G_F}{\sqrt{2}} V_{tb}^{\phantom{*}}V_{ts}^*\bigg(\sum_{k=7}^{10} C_k  \mathcal{O}_k+\sum_{i,j}C^{ij}_\nu \mathcal{O}^{ij}_\nu+ ...\bigg)\,, \ \ 
\end{equation}
where  the ellipsis denote  
four-quark operators,  $\mathcal{O}_7$ and $\mathcal{O}_8$ are
electro- and chromo-magnetic-moment-transition operators,  and $\mathcal{O}_9$, $\mathcal{O}_{10}$
and $\mathcal{O}_\nu$ are semi-leptonic operators involving either charged leptons or neutrinos:\footnote{We have assumed the $B$ anomalies in $R_{K^{(*)}}$ arise from a suppression in the $\mu$ channel relative to the SM. The Pati-Salam $(3,1)_{2/3}$ vector leptoquark model can equally well accommodate an enhancement in the electron channel, but a suppression in the muon channel is preferred by global fits to data that include angular moments in $B\to K^*\mu\mu$ and various branching fractions in addition to $R_{K^{(*)}}$.}
\begin{align}
\mathcal{O}_{9(10)}  &= 
    \frac{e^2}{(4 \pi)^2} \big[\bar s \gamma_\mu  P_ Lb\big]\big[\bar{\mu} \gamma^\mu (\gamma_5)\mu\big] \ , \\
    \mathcal{O}^{ij}_\nu &=\frac{2e^2}{(4 \pi)^2} \big[\bar s \gamma^\mu P_L b\big] \big[\bar \nu^i\gamma^\mu P_L\nu^j\big] \ .
\end{align}
Chirally-flipped $(b_{L(R)}\!\rightarrow \!b_{R(L)})$ versions
of all these operators are denoted by primes and are negligible in the SM.   New physics (NP) can generate modifications to the Wilson coefficients of the above operators, and, moreover, it can generate additional terms in the effective Lagrangian,   
\begin{equation}
\Delta\mathcal{L}=-
  \frac{4 G_F}{\sqrt{2}} V_{tb}^{\phantom{*}}V_{ts}^*\left(C_S  \mathcal{O}_S +C_P  \mathcal{O}_P + C'_S  \mathcal{O}'_S +C'_P  \mathcal{O}'_P  \right)
\end{equation}
in the form of scalar operators:
\begin{align}
  \mathcal{O}_S^{(\prime)} & = \frac{e^2}{(4 \pi)^2} \big[\bar s P_{R(L)} b\big] \big[\bar{\mu} \mu\big] \ , \\
  \mathcal{O}_P^{(\prime)}  &= \frac{e^2}{(4 \pi)^2} \big[\bar s P_{R(L)} b\big] \big[\bar{\mu} \gamma_5 \mu\big] \ .
\end{align}
Tensor operators cannot arise from short distance NP with the SM linearly realized and, moreover,
under this assumption $C_P=-C_S$ and $C'_P=C'_S$~\cite{Alonso:2014csa}. 

Exchange of the $(3,1)_{\frac23}$ vector leptoquark gives tree-level contributions to the Wilson coefficients
at its mass scale, $M$:
\begin{align}
\Delta C_9(M)&=-\Delta C_{10}(M) \!=\!-\frac{2\pi^2}{\sqrt2\, G_FM^2}\frac{g_4^2}{e^2}\frac{{L_{b\mu}^{d*}L_{s\mu}^{d}}}{V_{tb}^{\phantom{*}}V_{ts}^*} , \\
\Delta C'_9(M)&=\Delta C'_{10}(M)=-\frac{2\pi^2}{\sqrt2\,G_FM^2}\frac{g_4^2}{e^2}\frac{{R_{b\mu}^*R_{s\mu}}}{V_{tb}^{\phantom{*}}V_{ts}^*} \ ,\\
\Delta C_S(M)&=-\frac{4\pi^2}{\sqrt2\,G_FM^2}\frac{g_4^2}{e^2}\frac{{L^{d*}_{b\mu}R_{s\mu}}}{V_{tb}^{\phantom{*}}V_{ts}^*} \ ,\\
\Delta C'_S(M)&=-\frac{4\pi^2}{\sqrt2\,G_FM^2}\frac{g_4^2}{e^2}\frac{{R_{b\mu}^*L_{s\mu}^{d}}}{V_{tb}^{\phantom{*}}V_{ts}^*}\ ,\\
\Delta C^{ij}_\nu(M)&=0 \ .
\end{align}

The recent $B$ meson decay measurements  \cite{Aaij:2014ora,Aaij:2017vbb} show an excess above the SM background in the ratios $R_{K^{(*)}} =\Gamma(B\!\rightarrow \!K^{(*)}\mu\mu)\,/\,\Gamma(B\!\rightarrow\! K^{(*)}ee)$. 
Those anomalies are best fit by $\Delta C_9=-\Delta C_{10}\approx-0.6$ \cite{Geng:2017svp}, which requires $(g_4^2/M^2)L_{b\mu}^{d*}L_{s\mu}^{d}\approx 1.8\times 10^{-3}~\text{TeV}^{-2}$. Assuming $L_{b\mu}^{d*}L_{s\mu}^{d}=\tfrac12$, which is the largest value allowed by unitarity, we obtain the previously quoted leptoquark mass of $M \approx 16 \ {\rm TeV}$. Limits on $\Delta C'_{9,10}$ and $\Delta C_S^{(\prime)}$ can be accommodated by adjusting $R_{s\mu}$ and $R_{b\mu}$.

Experimental bounds on $R_{K^{(*)}\nu} =\text{Br}(B\to K^{(*)}\nu\bar\nu)/$ $\text{Br}(B\to
K^{(*)}\nu\bar\nu)^{\text{SM}}=\tfrac13\sum_{ij}|\delta^{ij}+C^{ij}_\nu/C_\nu^{\text{SM}}|^2$, where $C_\nu^{\text{SM}}\simeq-6.35$ \cite{Buras:2014fpa},  severely constrain theoretical models for those
anomalies.  As seen above, the $(3,1)_{\frac23}$ vector leptoquark evades this constraint by giving
no correction at all to $C_\nu$; the
result holds generally for this type of NP mediator at tree
level \cite{Alonso:2015sja,Calibbi:2015kma}. It has been pointed out that generally the condition 
$\Delta C_\nu(M)=0$ is not preserved by renormalization group running of the Wilson
coefficients \cite{Feruglio:2017rjo}. Because of the flavor structure of the interaction in Eq.~\eqref{eq:newINT} there are no ``penguin'' or wave function renormalization  contributions to  the running of $\Delta C_\nu$ down to the electroweak scale. The only contribution comes from the renormalization by exchange of SU(2) gauge bosons that mixes the singlet operator $(\bar q \gamma^\mu P_L e)(\bar e \gamma_\mu P_L q)$ into the triplet, $(\bar q\,\tau^a \gamma^\mu P_L e)(\bar e\,\tau^a \gamma_\mu P_L q)$, resulting in 
\begin{equation}
\Delta C^{ij}_\nu(M_W)=-\frac{3}{4\sqrt2 \,G_FM^2}\frac{g_4^2}{\sin^2\theta_w}\ln\left(\frac{M}{M_W}\right) \frac{S_{sj}S^*_{bi}}{V_{tb}^{\phantom{*}}V_{ts}^*} \ ,
\end{equation}
where $S\!=\!V^\dagger L^u\!=\!L^dU$. The vector contribution to the rate does not interfere with the SM, which implies $R_{K^{(*)}\nu} -1=$ $\frac13\sum_{ij} |C^{ij}_\nu/C_\nu^{\text{SM}}|^2$; using $R_{K^{(*)}\nu} <4.3$~\cite{Lutz:2013ftz}, we obtain the condition $M >0.8$~TeV.
Since $\ln(M/M_W)$ is not large for $M \!\approx\! 16 \ {\rm TeV}$, the leading log term is subject to sizable $(\approx 100\%)$ corrections. However, a complete one-loop calculation is beyond the scope of this work.

We pause to comment on the remarkable cancellation of the interference term between the SM and NP contributions to the rate for $B\to K^{(*)}\nu\bar\nu$ and the absence of a sum over generations in the pure NP contribution to the rate. These observations hold generally for any vector leptoquark model that couples universally to quark and lepton generations. This can be easily seen by not rotating to the neutrino mass eigenstate basis,  a good approximation for the nearly massless neutrinos. Vector  leptoquark exchange leads to an effective interaction with flavor structure $(\bar s_L \gamma^\mu\nu_L^2)(\bar \nu_L^3\gamma^\mu b_L)$, while the SM always involves a sum over the same neutrino flavors $\sim\sum_j \bar\nu_L^j\gamma^\mu\nu_L^j$. There are never common final states to the SM and the NP mediated interactions and therefore no interference. Moreover, there is a single flavor configuration in the final state of the NP mediated interaction ($\bar \nu^3 \nu^2$) while there are three configurations in the SM case ( $\bar \nu^j \nu^j$, $j=1,2,3$).

It has been suggested that the vector leptoquark $(3,1)_{\frac23}$ may alternatively be used to account for the anomaly in semileptonic 
decays to $\tau$-leptons~\cite{Sakaki:2013bfa,Alonso:2015sja}. Defining, as is customary,
$R_{D^{(*)}}=\text{Br}(B\to D^{(*)}\tau\nu)/\text{Br}(B\to D^{(*)}\ell\nu)$, the SM
predicts~\cite{Bernlochner:2017jka} (see also
\cite{Bigi:2017jbd,Jaiswal:2017rve,Fajfer:2012vx,Becirevic:2012jf}) $R_D=0.299(3)$ and
$R_{D^*}=0.257(3)$. These branching fractions have been measured by
Belle~\cite{Matyja:2007kt,Bozek:2010xy,Hirose:2016wfn}, BaBar~\cite{Lees:2012xj,Lees:2013uzd} and
LHCb~\cite{Aaij:2015yra}, and the average gives~\cite{Amhis:2016xyh} $R_D=0.403(47)$ and
$R_{D^*}=0.310(17)$. The effect of leptoquarks on  $B$ semileptonic decays to $\tau$ is described by
the following terms of the  effective Lagrangian for charged current interactions~\cite{Goldberger:1999yh,Cirigliano:2012ab}:\vspace{-1mm}
\begin{align}
\mathcal{L} \ \supset \ - \frac{4 G_F}{\sqrt{2}} \,V_{cb}\,&\Big[(U_{\tau j}+\epsilon^j_{L})\,(\bar c\gamma^\mu P_L b)(\bar \tau\gamma_\mu   P_L\nu^j)\nonumber\\
& + \epsilon_{s_R}^{j}\,(\bar c \,P_R\, b)(\bar \tau\, P_L\,\nu^j)\Big]+\text{h.c.}
\end{align}
 \vspace{-2mm}
 with
\begin{align} 
\!\!\!\epsilon_{L}^j \!=\!\frac{g_4^2}{4\sqrt2 \,G_FM^2}\frac{L^u_{cj}L^{d*}_{b\tau}}{V_{cb}} \ , \ \ \ \ 
\epsilon_{s_R}^j \!=\! - \frac{g_4^2}{2\sqrt2 \,G_FM^2}\frac{L^u_{cj}R^{*}_{b\tau}}{V_{cb}}.
\end{align} 
The $B_c$ lifetime~\cite{Alonso:2016oyd} and $B_c\to\tau\nu$ branching fraction~\cite{Akeroyd:2017mhr}
impose severe constraints on $\epsilon_{s_R}$; these are accommodated by abating
$R^{\phantom{u}}_{b\tau}$.  Hence, 
\begin{align}
\frac{R_{D^{(*)}}}{R_{D^{(*)}}^{\rm \, SM}} &\approx 1\!+\! 2\, {\rm Re}\Big[\sum_j\epsilon\, L^u_{cj} L^{d*}_{b\tau}\Big] \!\approx 1\!+\!2\, {\rm Re}\!\left[\epsilon\, (VL^d)_{c\tau} L^{d*}_{b\tau}\right]\nonumber\\
&\approx  1+ 2\, {\rm Re}\!\left[\epsilon\, L^d_{s\tau} L^{d*}_{b\tau}\right] \leq 1 + \epsilon \ ,
\end{align}
where,
\begin{equation}
\epsilon = \frac{g_4^2}{4\sqrt2 \,G_FV_{cb}M^2} \approx 0.1 \left(\frac{2 \ {\rm TeV}}{M}\right)^2  .
\end{equation}\vspace{0mm}

\subsection{Towards a viable UV completion}
We note that the simplest version of the model is heavily constrained by meson decay experiments. For generic, order one entries of the flavor matrices, the leptoquark mass is forced to be above the $1000 \ {\rm TeV}$ scale \cite{Valencia:1994cj,Smirnov:2007hv,Smirnov:2008zzb,Carpentier:2010ue,Kuznetsov:2012ai}. Surprisingly, all of the kaon decay bounds can be avoided if the unitary matrices $\hat{L}^d$ and $\hat{R}$ are of the form (see also \cite{Kuznetsov:2012ai}):
\bea\label{matricesL}
\hat{L}^d \approx \left( \  
\begin{matrix} \vspace{0.5mm}
0 & 0 & 1\\ \vspace{1mm}
L^d_{21} & L^d_{22} & 0\\ 
L^d_{31} & L^d_{32} &0
 \end{matrix} \ \ \right) , \ \ \  \ \hat{R} \approx  \left( \  
\begin{matrix} 
0 & 0 & 1\\ 
R_{21} & R_{22} & 0\\ 
R_{31} & R_{32} &0
 \end{matrix} \ \ \right), \ \ \ \ \ \ \ \ \eea
where it is actually sufficient that the entries labeled as zero are just $\lesssim 10^{-4}$.
Although current $\tau$ decay constraints are irrelevant for our choice of the leptoquark mass, with unsuppressed right-handed (RH) currents the $B$ meson decay bounds require $M \gtrsim 40 \ {\rm TeV}$ \cite{Kuznetsov:2012ai}.
Interestingly, we find that if a mechanism suppressing the RH currents is realized in nature, the present bounds from $B$ meson decays are much less stringent and require merely $M \gtrsim 19 \ {\rm TeV}$.

A possible way to suppress the RH currents is to associate them with a different gauge group than the left-handed (LH) ones, and to have the RH group spontaneously broken at a much higher scale than the LH group. A simple setting is offered by the gauge group
\bea
{\rm SU}(4)_L \times {\rm SU}(4)_R \times {\rm SU}(2)_L \times {\rm U}(1)
\eea
and does not require introducing any new fermion fields beyond the SM particle content and a RH neutrino:
\begin{equation} 
\label{eq:new2}
\begin{aligned}
(4,1, 2, 0) &= (3,2)_{\frac16} \oplus (1,2)_{-\frac12}\,,\\
(1,4,1, \tfrac12) &=   (3,1)_{\frac23} \oplus  (1,1)_0\,,\\
(1,4,1, -\tfrac12) &= (3,1)_{-\frac13} \oplus   (1,1)_{-1} \ .
\end{aligned}
\end{equation} 
Such a model predicts rates for  $B^+ \!\rightarrow\! K^+ e^\mp \mu^\pm$ and $\mu \rightarrow e\,\gamma$, among others, just above the experimental bounds reported in \cite{Aubert:2006vb,TheMEG:2016wtm}. The details of the model along with an analysis of the relevant experimental constraints will be the subject of a future publication \cite{work}.\footnote{Shortly after the results of our work became public, two other papers appeared proposing the explanation of $B$ decay anomalies by introducing new vector-like fields either within the Pati-Salam framework \cite{Calibbi:2017qbu} or in a model with an extended gauge group \cite{DiLuzio:2017vat}. }
 \vspace{1mm}

\section{Vector diquark model}
\label{other}
In this section we discuss the properties of a model with just one additional representation -- the vector color sextet:
\bea
 V_\mu = \left(\begin{matrix}
         V_u\\
         V_d \\
        \end{matrix}\right)_{\!\mu}^{\!\alpha\beta}= (6, 2)_{-\frac16} \ , 
 \eea 
which is  obviously  free from proton decay. Although in the SM all fundamental vector particles are gauge bosons, we can still imagine that such a vector diquark
 arises from a vector GUT representation, for instance from a vector $40$ irrep of ${\rm SU}(5)$
 \cite{Slansky:1981yr}. 
The Lagrangian for the model  is given by:
\bea\label{Lag1}
\mathcal{L}_{V} &=&-\tfrac{1}{4} (D_{[\mu} V_{\nu]})^\dagger D^{[\mu} V^{\nu]}+M^2V_\mu^\dagger V^\mu \nonumber\\
&& -   \left[ \lambda_{ij} \,(\overline{Q}^c_{L})_{\alpha}^i \gamma^\mu (d_{R})_{\beta}^j (V^\dagger)^{\alpha\beta}_{\mu} + {\rm h.c.} \right]  , \ \ \ \ 
\eea
where $\alpha, \beta = 1, 2, 3 $ are ${\rm SU}(3)_c$ indices, $i, j = 1, 2, 3$ are family
indices and there is an implicit contraction of the ${\rm SU}(2)_L$ indices. We assume that the  mass
term  arises from  a consistent UV completion.
 
 Among the allowed higher dimensional operators, $n-\bar{n}$ oscillations, as we discuss in Sec.~\ref{nn}, are mediated by the dimension five terms:
 \bea\label{o1o2}
 \begin{aligned}
\mathcal{O}_1 &=  \frac{c_1}{\Lambda}\, V_\mu^{\alpha \alpha'} \epsilon \, V_\nu^{\beta \beta'}   ( \bar u^c_R)^\delta \sigma^{\mu\nu} d_R^{\delta'} \, \epsilon_{\alpha \beta \delta}  \,\epsilon_{\alpha'\beta'\delta'} \ ,\\
\mathcal{O}_2 &=  \frac{c_2}{\Lambda}\, \big[\partial_\mu (V^\mu)^{\alpha \alpha'} \epsilon \,V_{\nu}^{\beta \beta'} \big]\big[(V^\nu)^{\delta \delta'} \epsilon \, H\big]  \epsilon_{\alpha \beta \delta}  \,\epsilon_{\alpha'\beta'\delta'} \ .
\end{aligned}
 \eea

\subsection{LHC bounds}
\label{sec:LHC}

Several studies of constraints and prospects for discovering vector diquarks at the
LHC can be found in the literature. Most of the
analyses have focused on the case of a sizable diquark coupling to quarks
\cite{Arik:2001bc,Sahin:2009dca,Zhang:2010kr}, although an LHC four-jet search that is essentially independent of the strength of the diquark coupling to quarks has also been considered \cite{Richardson:2011df}.

There are severe limits on vector diquark masses arising from LHC searches for non-SM dijet signals
\cite{Sirunyan:2016iap,Aaboud:2017yvp}. For a coupling $\lambda_{ij} \approx 1$ $(i,j=1,2)$ those
searches result in a bound on the vector diquark mass
\bea
M_{\lambda \approx 1}\gtrsim 8 \ {\rm TeV} \ .
\eea


Lowering the value of the coupling  to  $\lambda_{ij} \approx 0.01$ completely removes the LHC constraints from dijet searches and, at the same time, does not affect the strength of the four-jet signal arising from gluon fusion. Using the results of the analysis of four-jet events at the LHC presented in \cite{Richardson:2011df}, the currently collected  $37 \ {\rm fb}^{-1}$ of data by the ATLAS experiment \cite{Aaboud:2017yvp}
 with no evident excess above the SM background constrains the vector diquark mass to be
\bea\label{m1}
M_{\lambda\ll 1} \gtrsim 2.5 \ {\rm TeV} \ .
\eea

Additional processes constraining the vector diquark model include neutral meson mixing and radiative $B$ meson decays \cite{Isidori:2010kg,Blake:2016olu}. The resulting bounds  in the case of scalar diquark models were calculated in \cite{Arnold:2012sd,Maalampi:1987pb,Chakraverty:2000rm} and are similar here.

\subsection{Neutron-antineutron oscillations}
\label{nn}
In the light of null  results from proton decay searches \cite{Miura:2016krn},  the possibility of
discovering  $n - \bar{n}$ oscillations has recently gained increased interest
\cite{Phillips:2014fgb}.  The most important reason for this is that the matter-antimatter asymmetry
of the Universe requires baryon number to be violated at some point during its evolution. If
processes with $\Delta B = 1$ are indeed  suppressed or do not occur in Nature at all, the next
simplest case involves $\Delta B \!=\! 2$,  a baryon number
breaking pattern  that may result in  $n - \bar{n}$ oscillations without
  proton decay. 
Moreover, the SM augmented only by 
RH neutrinos can have an  additional ${\rm U}(1)_{B-L}$ gauge symmetry
without introducing any gauge anomalies. Through this symmetry, processes with $\Delta B = 2$ would
be accompanied by $\Delta L \!=\! 2$ lepton number violating ones, which in turn are  intrinsically connected to the seesaw mechanism generating naturally small neutrino masses \cite{Minkowski:1977sc}. 

Models constructed so far propose $n - \bar{n}$ oscillations mediated by scalar diquarks, as
mentioned in Sec.~\ref{dq}. Those oscillations proceed through a triple scalar vertex,
 so that  the process is described at low energies by a
  local operator of dimension nine.  Below we show  that $n - \bar{n}$ oscillations can also be
mediated by vector diquarks, in particular within the model with just one new representation
discussed in this section. 

\begin{figure}[t!]
\includegraphics[width=0.75\linewidth]{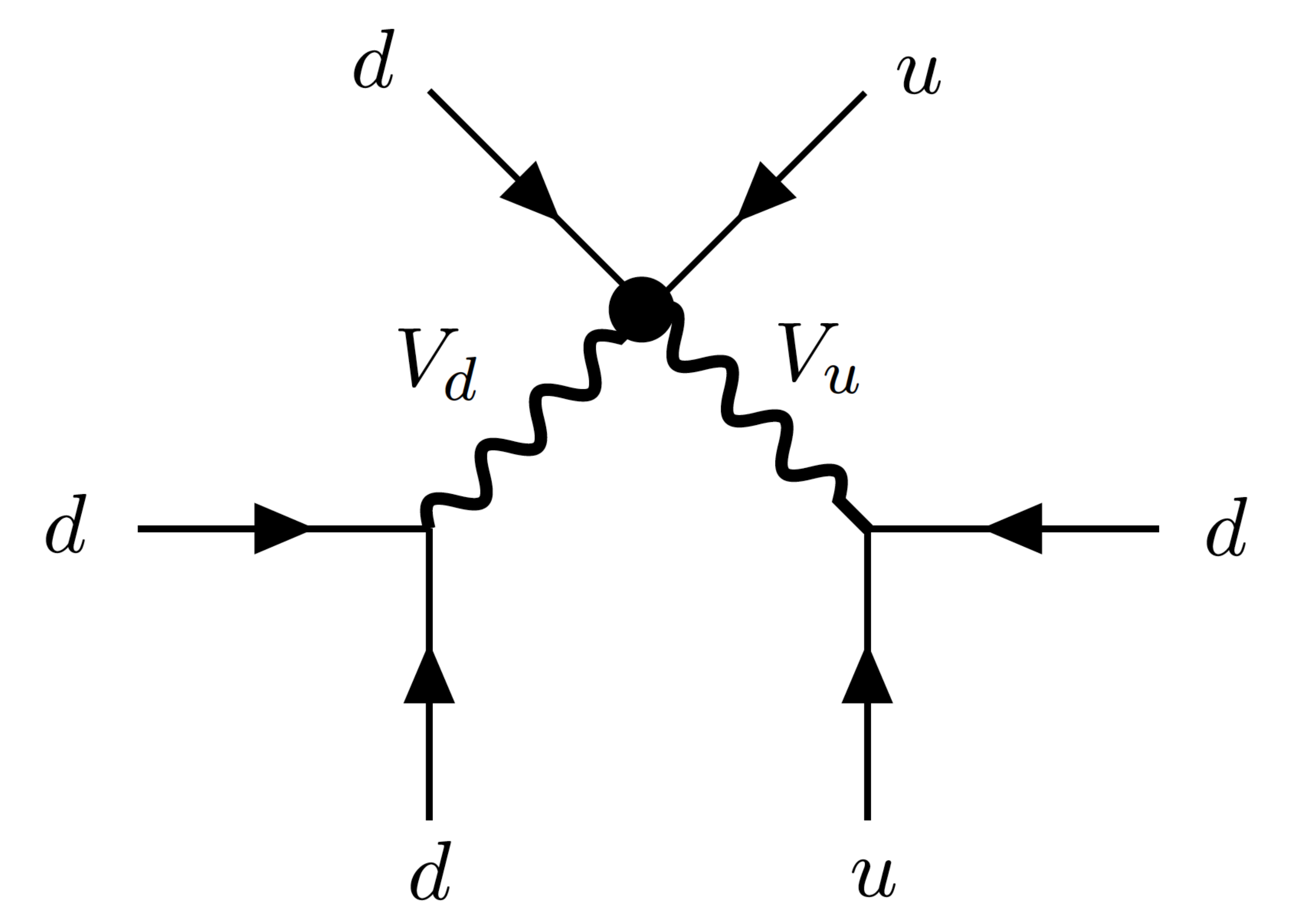} \vspace{-1mm}
\caption{
\small{Process mediating $n \!-\! \bar{n}$ oscillations via the vector diquark \\ {\centerline{$V = (6,2)_{-\frac16}$.}}}}
\label{fig:4}
\end{figure}

The least suppressed channel is via a dimension five gauge invariant quartic interaction $\mathcal{O}_1$ in Eq.~(\ref{o1o2})
involving two vector diquarks $(6,2)_{-\frac16}$ and the SM up and down quarks, as shown in Fig.~\ref{fig:4},
ultimately leading to $n - \bar{n}$ oscillations  through a low energy  effective interaction
 local operator of dimension nine, as in the case of scalar diquarks.\footnote{We note that $n - \bar{n}$ oscillations occur at the renormalizable level in a model involving the vector diquarks $(6,2)_{-1/6}$, $(6,2)_{5/6}$, and the scalar diquark  $(6,1)_{-2/3}$, and proceed through a triple sextet interaction.\vspace{1mm} }

With the simplifying assumption that $c_1 \!\approx\! 1$ and neglecting other operators contributing to the signal we now estimate  the rate of  $n - \bar{n}$ oscillations in this model. 
The  effective  Hamiltonian corresponding to the operator $\mathcal{O}_1$ is:
\bea
&&\mathcal{H_{\rm eff}} \approx  -\frac{\lambda_{11}^2 }{M^4 \Lambda} (\overline{d}^c_{L})^{{\alpha}}\gamma_\mu d_{R}^{\alpha'}  (\overline{u}^c_{L})^{{\beta}}\gamma_\nu d_{R}^{\beta'}  (\overline{u}^c_{R})^{{\delta}}\sigma^{\mu\nu} d_{R}^{\delta'} \nonumber\\
& &\times \left(\epsilon_{\alpha\beta\delta} \,\epsilon_{\alpha'\beta'\delta'} + \epsilon_{\alpha'\beta\delta} \,\epsilon_{\alpha\beta'\delta'}+\epsilon_{\alpha\beta'\delta} \,\epsilon_{\alpha'\beta\delta'}+\epsilon_{\alpha\beta\delta'} \,\epsilon_{\alpha'\beta'\delta}\right) \nonumber\\
& & + \ {\rm h.c.} \ .
\eea
Combining this with the results of \cite{Arnold:2012sd,Tsutsui:2004qc}, we obtain an estimate for the $n - \bar{n}$ transition matrix element,\footnote{The single particle state normalization adopted here is:
\bea
\langle n(\vec{p}\,) | n(\vec{p}\,') \rangle = \frac{E}{m}(2\pi)^3\delta^{(3)}(\vec{p}-\vec{p}\,') \ .
\eea}
\bea
\big|\langle \bar{n}| \mathcal{H}_{\rm eff} | n \rangle \big|  \approx  \frac{10^{-4}\,|\lambda_{11}^2|}{M^4 \Lambda}  \ {\rm GeV^6} \ ,
\eea
where $|n\rangle$ is the neutron state at zero momentum.
Current experimental limit \cite{Abe:2011ky} implies, assuming $\lambda_{11} \approx 0.01$ (which is well below the LHC bound from dijet searches, as discussed earlier), that
\bea\label{lim}
M \gtrsim 2.5\ {\rm TeV} \left(\frac{10^8\ {\rm TeV}}{\Lambda}\right)^{1/4}  .
\eea

An interesting limit on the vector diquark mass is derived if we assume that the physics behind the triple diquark interaction with the SM Higgs is related to the physics responsible for providing the diquark its mass, i.e., for  $\Lambda \approx M$. In such case the $n - \bar{n}$ oscillation search provides the bound:
\bea
M \gtrsim 90 \ {\rm TeV} \, ,
\eea
much stronger than the LHC limit. 
If there is new physics around that energy scale, it should be discovered by future $n - \bar{n}$ oscillation experiments with increased sensitivity \cite{Phillips:2014fgb}, which are going to probe  the vector diquark mass scale up to  \, $\sim 175 \ {\rm TeV}$.  This is especially interesting since models with TeV-scale diquarks tend to improve gauge coupling unification \cite{Babu:2012vc,Babu:2013yca}.

\section{Conclusions}
\label{conclusions} 

We have shown  that lack of experimental evidence for proton decay singles out only a handful of phenomenologically viable leptoquark  models. In addition, even leptoquark models with tree-level proton stability contain dangerous dimension five proton decay mediating operators and require an appropriate UV completion to remain consistent with experiments. This is especially relevant  for the Standard Model extension involving the  vector leptoquarks $(3,1)_{\frac23}$ or $(3,3)_{\frac23}$, since those are the only two models with a single new representation that do not suffer from tree-level proton decay and can explain the recently discovered anomalies in $B$ meson decays. 

The property which makes the vector leptoquark $(3,1)_{\frac23}$ even more appealing is that it fits perfectly into the simplest Pati-Salam unification model, where it can be identified as the new gauge boson. If such an exciting  scenario is indeed realized in nature, the $B$ physics experiments can be used to actually probe the scale and various properties of grand unification!

In the second part of the paper we focused on a model with a vector diquark $(6,2)_{-\frac16}$ and showed that neutron-antineutron oscillations can be mediated by such a vector particle. The model is somewhat constrained by LHC dijet  searches; however, it can still yield a sizable neutron-antineutron oscillation signal,  that can be probed in current and upcoming experiments. It can also  give rise to significant four-jet event rates testable  at the LHC. 

It would be interesting to explore whether a vector diquark with a mass at the TeV scale can improve gauge coupling unification in non-supersymmetric grand unified theories, similarly to the scalar case \cite{Stone:2011dn}, providing  even more motivation for upgrading the neutron-antineutron oscillation experimental sensitivity.

\subsection*{Acknowledgments}

We thank Aneesh Manohar for useful conversations  and Pavel Fileviez P\'{e}rez for comments regarding the manuscript. This research was supported in part by the DOE Grant No.~${\rm DE}$-{\rm SC}0009919.

\bibliography{bibliography}

\begin{thebibliography}{103}%
\makeatletter
\providecommand \@ifxundefined [1]{%
 \@ifx{#1\undefined}
}%
\providecommand \@ifnum [1]{%
 \ifnum #1\expandafter \@firstoftwo
 \else \expandafter \@secondoftwo
 \fi
}%
\providecommand \@ifx [1]{%
 \ifx #1\expandafter \@firstoftwo
 \else \expandafter \@secondoftwo
 \fi
}%
\providecommand \natexlab [1]{#1}%
\providecommand \enquote  [1]{``#1''}%
\providecommand \bibnamefont  [1]{#1}%
\providecommand \bibfnamefont [1]{#1}%
\providecommand \citenamefont [1]{#1}%
\providecommand \href@noop [0]{\@secondoftwo}%
\providecommand \href [0]{\begingroup \@sanitize@url \@href}%
\providecommand \@href[1]{\@@startlink{#1}\@@href}%
\providecommand \@@href[1]{\endgroup#1\@@endlink}%
\providecommand \@sanitize@url [0]{\catcode `\\12\catcode `\$12\catcode
  `\&12\catcode `\#12\catcode `\^12\catcode `\_12\catcode `\%12\relax}%
\providecommand \@@startlink[1]{}%
\providecommand \@@endlink[0]{}%
\providecommand \url  [0]{\begingroup\@sanitize@url \@url }%
\providecommand \@url [1]{\endgroup\@href {#1}{\urlprefix }}%
\providecommand \urlprefix  [0]{URL }%
\providecommand \Eprint [0]{\href }%
\providecommand \doibase [0]{http://dx.doi.org/}%
\providecommand \selectlanguage [0]{\@gobble}%
\providecommand \bibinfo  [0]{\@secondoftwo}%
\providecommand \bibfield  [0]{\@secondoftwo}%
\providecommand \translation [1]{[#1]}%
\providecommand \BibitemOpen [0]{}%
\providecommand \bibitemStop [0]{}%
\providecommand \bibitemNoStop [0]{.\EOS\space}%
\providecommand \EOS [0]{\spacefactor3000\relax}%
\providecommand \BibitemShut  [1]{\csname bibitem#1\endcsname}%
\let\auto@bib@innerbib\@empty
\bibitem [{\citenamefont {Georgi}\ and\ \citenamefont
  {Glashow}(1974)}]{Georgi:1974sy}%
  \BibitemOpen
  \bibfield  {author} {\bibinfo {author} {\bibfnamefont {H.}~\bibnamefont
  {Georgi}}\ and\ \bibinfo {author} {\bibfnamefont {S.~L.}\ \bibnamefont
  {Glashow}},\ }\bibfield  {title} {\enquote {\bibinfo {title} {\emph{Unity of
  all Elementary Particle Forces}},}\ }\href {\doibase
  10.1103/PhysRevLett.32.438} {\bibfield  {journal} {\bibinfo  {journal} {Phys.
  Rev. Lett.}\ }\textbf {\bibinfo {volume} {32}},\ \bibinfo {pages} {438--441}
  (\bibinfo {year} {1974})}\BibitemShut {NoStop}%
\bibitem [{\citenamefont {Fritzsch}\ and\ \citenamefont
  {Minkowski}(1975)}]{Fritzsch:1974nn}%
  \BibitemOpen
  \bibfield  {author} {\bibinfo {author} {\bibfnamefont {H.}~\bibnamefont
  {Fritzsch}}\ and\ \bibinfo {author} {\bibfnamefont {P.}~\bibnamefont
  {Minkowski}},\ }\bibfield  {title} {\enquote {\bibinfo {title} {\emph{Unified
  Interactions of Leptons and Hadrons}},}\ }\href {\doibase
  10.1016/0003-4916(75)90211-0} {\bibfield  {journal} {\bibinfo  {journal}
  {Annals Phys.}\ }\textbf {\bibinfo {volume} {93}},\ \bibinfo {pages}
  {193--266} (\bibinfo {year} {1975})}\BibitemShut {NoStop}%
\bibitem [{\citenamefont {Karananas}\ and\ \citenamefont
  {Shaposhnikov}(2017)}]{Karananas:2017mxm}%
  \BibitemOpen
  \bibfield  {author} {\bibinfo {author} {\bibfnamefont {G.~K.}\ \bibnamefont
  {Karananas}}\ and\ \bibinfo {author} {\bibfnamefont {M.}~\bibnamefont
  {Shaposhnikov}},\ }\bibfield  {title} {\enquote {\bibinfo {title}
  {\emph{Gauge Coupling Unification without Leptoquarks}},}\ }\href {\doibase
  10.1016/j.physletb.2017.05.065} {\bibfield  {journal} {\bibinfo  {journal}
  {Phys. Lett.}\ }\textbf {\bibinfo {volume} {B771}},\ \bibinfo {pages}
  {332--338} (\bibinfo {year} {2017})},\ \Eprint
  {http://arxiv.org/abs/1703.02964} {arXiv:1703.02964 [hep-ph]} \BibitemShut
  {NoStop}%
\bibitem [{\citenamefont {Fornal}\ and\ \citenamefont
  {Grinstein}(2017)}]{Fornal:2017xcj}%
  \BibitemOpen
  \bibfield  {author} {\bibinfo {author} {\bibfnamefont {B.}~\bibnamefont
  {Fornal}}\ and\ \bibinfo {author} {\bibfnamefont {B.}~\bibnamefont
  {Grinstein}},\ }\bibfield  {title} {\enquote {\bibinfo {title} {\emph{SU(5)
  Unification without Proton Decay}},}\ }\href@noop {} {\bibfield  {journal}
  {\bibinfo  {journal} {accepted by Phys. Rev. Lett.}\ } (\bibinfo {year}
  {2017})},\ \Eprint {http://arxiv.org/abs/1706.08535} {arXiv:1706.08535
  [hep-ph]} \BibitemShut {NoStop}%
\bibitem [{\citenamefont {Nath}\ and\ \citenamefont
  {Fileviez~Perez}(2007)}]{Nath:2006ut}%
  \BibitemOpen
  \bibfield  {author} {\bibinfo {author} {\bibfnamefont {P.}~\bibnamefont
  {Nath}}\ and\ \bibinfo {author} {\bibfnamefont {P.}~\bibnamefont
  {Fileviez~Perez}},\ }\bibfield  {title} {\enquote {\bibinfo {title}
  {\emph{Proton Stability in Grand Unified Theories, in Strings and in
  Branes}},}\ }\href {\doibase 10.1016/j.physrep.2007.02.010} {\bibfield
  {journal} {\bibinfo  {journal} {Phys. Rept.}\ }\textbf {\bibinfo {volume}
  {441}},\ \bibinfo {pages} {191--317} (\bibinfo {year} {2007})},\ \Eprint
  {http://arxiv.org/abs/hep-ph/0601023} {arXiv:hep-ph/0601023 [hep-ph]}
  \BibitemShut {NoStop}%
\bibitem [{\citenamefont {Abe}\ \emph {et~al.}(2017)\citenamefont {Abe} \emph
  {et~al.}}]{Miura:2016krn}%
  \BibitemOpen
  \bibfield  {author} {\bibinfo {author} {\bibfnamefont {K.}~\bibnamefont
  {Abe}} \emph {et~al.} (\bibinfo {collaboration} {Super-Kamiokande}),\
  }\bibfield  {title} {\enquote {\bibinfo {title} {\emph{{Search for Proton
  Decay via $p \!\to \!e^+\pi^0$ and $p \!\to\! \mu^+\pi^0$ in 0.31
  Megaton$\cdot$years Exposure of the Super-Kamiokande Water Cherenkov
  Detector}}},}\ }\href {\doibase 10.1103/PhysRevD.95.012004} {\bibfield
  {journal} {\bibinfo  {journal} {Phys. Rev.}\ }\textbf {\bibinfo {volume}
  {D95}},\ \bibinfo {pages} {012004} (\bibinfo {year} {2017})},\ \Eprint
  {http://arxiv.org/abs/1610.03597} {arXiv:1610.03597 [hep-ex]} \BibitemShut
  {NoStop}%
\bibitem [{\citenamefont {Arnold}\ \emph
  {et~al.}(2013{\natexlab{a}})\citenamefont {Arnold}, \citenamefont {Fornal},\
  and\ \citenamefont {Wise}}]{Arnold:2012sd}%
  \BibitemOpen
  \bibfield  {author} {\bibinfo {author} {\bibfnamefont {J.~M.}\ \bibnamefont
  {Arnold}}, \bibinfo {author} {\bibfnamefont {B.}~\bibnamefont {Fornal}}, \
  and\ \bibinfo {author} {\bibfnamefont {M.~B.}\ \bibnamefont {Wise}},\
  }\bibfield  {title} {\enquote {\bibinfo {title} {\emph{Simplified Models with
  Baryon Number Violation but No Proton Decay}},}\ }\href {\doibase
  10.1103/PhysRevD.87.075004} {\bibfield  {journal} {\bibinfo  {journal} {Phys.
  Rev.}\ }\textbf {\bibinfo {volume} {D87}},\ \bibinfo {pages} {075004}
  (\bibinfo {year} {2013}{\natexlab{a}})},\ \Eprint
  {http://arxiv.org/abs/1212.4556} {arXiv:1212.4556 [hep-ph]} \BibitemShut
  {NoStop}%
\bibitem [{\citenamefont {Arnold}\ \emph
  {et~al.}(2013{\natexlab{b}})\citenamefont {Arnold}, \citenamefont {Fornal},\
  and\ \citenamefont {Wise}}]{Arnold:2013cva}%
  \BibitemOpen
  \bibfield  {author} {\bibinfo {author} {\bibfnamefont {J.~M.}\ \bibnamefont
  {Arnold}}, \bibinfo {author} {\bibfnamefont {B.}~\bibnamefont {Fornal}}, \
  and\ \bibinfo {author} {\bibfnamefont {M.~B.}\ \bibnamefont {Wise}},\
  }\bibfield  {title} {\enquote {\bibinfo {title} {\emph{Phenomenology of
  Scalar Leptoquarks}},}\ }\href {\doibase 10.1103/PhysRevD.88.035009}
  {\bibfield  {journal} {\bibinfo  {journal} {Phys. Rev.}\ }\textbf {\bibinfo
  {volume} {D88}},\ \bibinfo {pages} {035009} (\bibinfo {year}
  {2013}{\natexlab{b}})},\ \Eprint {http://arxiv.org/abs/1304.6119}
  {arXiv:1304.6119 [hep-ph]} \BibitemShut {NoStop}%
\bibitem [{\citenamefont {Aaij}\ \emph {et~al.}(2014)\citenamefont {Aaij} \emph
  {et~al.}}]{Aaij:2014ora}%
  \BibitemOpen
  \bibfield  {author} {\bibinfo {author} {\bibfnamefont {R.}~\bibnamefont
  {Aaij}} \emph {et~al.} (\bibinfo {collaboration} {LHCb}),\ }\bibfield
  {title} {\enquote {\bibinfo {title} {\emph{Test of Lepton Universality Using
  $B^{+}\!\rightarrow\!K^{+}\ell^{+}\ell^{-}$ Decays}},}\ }\href {\doibase
  10.1103/PhysRevLett.113.151601} {\bibfield  {journal} {\bibinfo  {journal}
  {Phys. Rev. Lett.}\ }\textbf {\bibinfo {volume} {113}},\ \bibinfo {pages}
  {151601} (\bibinfo {year} {2014})},\ \Eprint {http://arxiv.org/abs/1406.6482}
  {arXiv:1406.6482 [hep-ex]} \BibitemShut {NoStop}%
\bibitem [{\citenamefont {Aaij}\ \emph {et~al.}(2017)\citenamefont {Aaij} \emph
  {et~al.}}]{Aaij:2017vbb}%
  \BibitemOpen
  \bibfield  {author} {\bibinfo {author} {\bibfnamefont {R.}~\bibnamefont
  {Aaij}} \emph {et~al.} (\bibinfo {collaboration} {LHCb}),\ }\bibfield
  {title} {\enquote {\bibinfo {title} {\emph{Test of Lepton Universality with
  $B^{0} \rightarrow K^{*0}\ell^{+}\ell^{-}$ Decays}},}\ }\href {\doibase
  10.1007/JHEP08(2017)055} {\bibfield  {journal} {\bibinfo  {journal} {JHEP}\
  }\textbf {\bibinfo {volume} {08}},\ \bibinfo {pages} {055} (\bibinfo {year}
  {2017})},\ \Eprint {http://arxiv.org/abs/1705.05802} {arXiv:1705.05802
  [hep-ex]} \BibitemShut {NoStop}%
\bibitem [{\citenamefont {Alonso}\ \emph {et~al.}(2015)\citenamefont {Alonso},
  \citenamefont {Grinstein},\ and\ \citenamefont
  {Martin~Camalich}}]{Alonso:2015sja}%
  \BibitemOpen
  \bibfield  {author} {\bibinfo {author} {\bibfnamefont {R.}~\bibnamefont
  {Alonso}}, \bibinfo {author} {\bibfnamefont {B.}~\bibnamefont {Grinstein}}, \
  and\ \bibinfo {author} {\bibfnamefont {J.}~\bibnamefont {Martin~Camalich}},\
  }\bibfield  {title} {\enquote {\bibinfo {title} {\emph{Lepton Universality
  Violation and Lepton Flavor Conservation in $B$-meson Decays}},}\ }\href
  {\doibase 10.1007/JHEP10(2015)184} {\bibfield  {journal} {\bibinfo  {journal}
  {JHEP}\ }\textbf {\bibinfo {volume} {10}},\ \bibinfo {pages} {184} (\bibinfo
  {year} {2015})},\ \Eprint {http://arxiv.org/abs/1505.05164} {arXiv:1505.05164
  [hep-ph]} \BibitemShut {NoStop}%
\bibitem [{\citenamefont {Dorsner}\ \emph {et~al.}(2016)\citenamefont
  {Dorsner}, \citenamefont {Fajfer}, \citenamefont {Greljo}, \citenamefont
  {Kamenik},\ and\ \citenamefont {Kosnik}}]{Dorsner:2016wpm}%
  \BibitemOpen
  \bibfield  {author} {\bibinfo {author} {\bibfnamefont {I.}~\bibnamefont
  {Dorsner}}, \bibinfo {author} {\bibfnamefont {S.}~\bibnamefont {Fajfer}},
  \bibinfo {author} {\bibfnamefont {A.}~\bibnamefont {Greljo}}, \bibinfo
  {author} {\bibfnamefont {J.~F.}\ \bibnamefont {Kamenik}}, \ and\ \bibinfo
  {author} {\bibfnamefont {N.}~\bibnamefont {Kosnik}},\ }\bibfield  {title}
  {\enquote {\bibinfo {title} {\emph{Physics of Leptoquarks in Precision
  Experiments and at Particle Colliders}},}\ }\href {\doibase
  10.1016/j.physrep.2016.06.001} {\bibfield  {journal} {\bibinfo  {journal}
  {Phys. Rept.}\ }\textbf {\bibinfo {volume} {641}},\ \bibinfo {pages} {1--68}
  (\bibinfo {year} {2016})},\ \Eprint {http://arxiv.org/abs/1603.04993}
  {arXiv:1603.04993 [hep-ph]} \BibitemShut {NoStop}%
\bibitem [{\citenamefont {Patrignani}\ \emph {et~al.}(2016)\citenamefont
  {Patrignani} \emph {et~al.}}]{Olive:2016xmw}%
  \BibitemOpen
  \bibfield  {author} {\bibinfo {author} {\bibfnamefont {C.}~\bibnamefont
  {Patrignani}} \emph {et~al.} (\bibinfo {collaboration} {Particle Data
  Group}),\ }\bibfield  {title} {\enquote {\bibinfo {title} {\emph{Review of
  Particle Physics}},}\ }\href {\doibase 10.1088/1674-1137/40/10/100001}
  {\bibfield  {journal} {\bibinfo  {journal} {Chin. Phys.}\ }\textbf {\bibinfo
  {volume} {C40}},\ \bibinfo {pages} {100001} (\bibinfo {year}
  {2016})}\BibitemShut {NoStop}%
\bibitem [{\citenamefont {Buchmuller}\ \emph {et~al.}(1987)\citenamefont
  {Buchmuller}, \citenamefont {Ruckl},\ and\ \citenamefont
  {Wyler}}]{Buchmuller:1986zs}%
  \BibitemOpen
  \bibfield  {author} {\bibinfo {author} {\bibfnamefont {W.}~\bibnamefont
  {Buchmuller}}, \bibinfo {author} {\bibfnamefont {R.}~\bibnamefont {Ruckl}}, \
  and\ \bibinfo {author} {\bibfnamefont {D.}~\bibnamefont {Wyler}},\ }\bibfield
   {title} {\enquote {\bibinfo {title} {\emph{Leptoquarks in Lepton-Quark
  Collisions}},}\ }\href {\doibase 10.1016/S0370-2693(99)00014-3,
  10.1016/0370-2693(87)90637-X} {\bibfield  {journal} {\bibinfo  {journal}
  {Phys. Lett.}\ }\textbf {\bibinfo {volume} {B191}},\ \bibinfo {pages}
  {442--448} (\bibinfo {year} {1987})},\ \bibinfo {note} {[Erratum: Phys.
  Lett.B448,320(1999)]}\BibitemShut {NoStop}%
\bibitem [{\citenamefont {Davidson}\ \emph {et~al.}(1994)\citenamefont
  {Davidson}, \citenamefont {Bailey},\ and\ \citenamefont
  {Campbell}}]{Davidson:1993qk}%
  \BibitemOpen
  \bibfield  {author} {\bibinfo {author} {\bibfnamefont {S.}~\bibnamefont
  {Davidson}}, \bibinfo {author} {\bibfnamefont {D.~C.}\ \bibnamefont
  {Bailey}}, \ and\ \bibinfo {author} {\bibfnamefont {B.~A.}\ \bibnamefont
  {Campbell}},\ }\bibfield  {title} {\enquote {\bibinfo {title} {\emph{Model
  Independent Constraints on Leptoquarks from Rare Processes}},}\ }\href
  {\doibase 10.1007/BF01552629} {\bibfield  {journal} {\bibinfo  {journal} {Z.
  Phys.}\ }\textbf {\bibinfo {volume} {C61}},\ \bibinfo {pages} {613--644}
  (\bibinfo {year} {1994})},\ \Eprint {http://arxiv.org/abs/hep-ph/9309310}
  {arXiv:hep-ph/9309310 [hep-ph]} \BibitemShut {NoStop}%
\bibitem [{\citenamefont {Hewett}\ and\ \citenamefont
  {Rizzo}(1997)}]{Hewett:1997ce}%
  \BibitemOpen
  \bibfield  {author} {\bibinfo {author} {\bibfnamefont {J.~L.}\ \bibnamefont
  {Hewett}}\ and\ \bibinfo {author} {\bibfnamefont {T.~G.}\ \bibnamefont
  {Rizzo}},\ }\bibfield  {title} {\enquote {\bibinfo {title} {\emph{Much Ado
  About Leptoquarks: A Comprehensive Analysis}},}\ }\href {\doibase
  10.1103/PhysRevD.56.5709} {\bibfield  {journal} {\bibinfo  {journal} {Phys.
  Rev.}\ }\textbf {\bibinfo {volume} {D56}},\ \bibinfo {pages} {5709--5724}
  (\bibinfo {year} {1997})},\ \Eprint {http://arxiv.org/abs/hep-ph/9703337}
  {arXiv:hep-ph/9703337 [hep-ph]} \BibitemShut {NoStop}%
\bibitem [{\citenamefont {Geng}\ \emph {et~al.}(2017)\citenamefont {Geng},
  \citenamefont {Grinstein}, \citenamefont {Jager}, \citenamefont
  {Martin~Camalich}, \citenamefont {Ren},\ and\ \citenamefont
  {Shi}}]{Geng:2017svp}%
  \BibitemOpen
  \bibfield  {author} {\bibinfo {author} {\bibfnamefont {L.-S.}\ \bibnamefont
  {Geng}}, \bibinfo {author} {\bibfnamefont {B.}~\bibnamefont {Grinstein}},
  \bibinfo {author} {\bibfnamefont {S.}~\bibnamefont {Jager}}, \bibinfo
  {author} {\bibfnamefont {J.}~\bibnamefont {Martin~Camalich}}, \bibinfo
  {author} {\bibfnamefont {X.-L.}\ \bibnamefont {Ren}}, \ and\ \bibinfo
  {author} {\bibfnamefont {R.-X.}\ \bibnamefont {Shi}},\ }\bibfield  {title}
  {\enquote {\bibinfo {title} {\emph{Towards the Discovery of New Physics with
  Lepton-Universality Ratios of $b\to s\ell\ell$ Decays}},}\ }\href {\doibase
  10.1103/PhysRevD.96.093006} {\bibfield  {journal} {\bibinfo  {journal} {Phys.
  Rev.}\ }\textbf {\bibinfo {volume} {D96}},\ \bibinfo {pages} {093006}
  (\bibinfo {year} {2017})},\ \Eprint {http://arxiv.org/abs/1704.05446}
  {arXiv:1704.05446 [hep-ph]} \BibitemShut {NoStop}%
\bibitem [{\citenamefont {Ciuchini}\ \emph {et~al.}(2017)\citenamefont
  {Ciuchini}, \citenamefont {Coutinho}, \citenamefont {Fedele}, \citenamefont
  {Franco}, \citenamefont {Paul}, \citenamefont {Silvestrini},\ and\
  \citenamefont {Valli}}]{Ciuchini:2017mik}%
  \BibitemOpen
  \bibfield  {author} {\bibinfo {author} {\bibfnamefont {M.}~\bibnamefont
  {Ciuchini}}, \bibinfo {author} {\bibfnamefont {A.~M.}\ \bibnamefont
  {Coutinho}}, \bibinfo {author} {\bibfnamefont {M.}~\bibnamefont {Fedele}},
  \bibinfo {author} {\bibfnamefont {E.}~\bibnamefont {Franco}}, \bibinfo
  {author} {\bibfnamefont {A.}~\bibnamefont {Paul}}, \bibinfo {author}
  {\bibfnamefont {L.}~\bibnamefont {Silvestrini}}, \ and\ \bibinfo {author}
  {\bibfnamefont {M.}~\bibnamefont {Valli}},\ }\bibfield  {title} {\enquote
  {\bibinfo {title} {\emph{On Flavourful Easter Eggs for New Physics Hunger and
  Lepton Flavour Universality Violation}},}\ }\href {\doibase
  10.1140/epjc/s10052-017-5270-2} {\bibfield  {journal} {\bibinfo  {journal}
  {Eur. Phys. J.}\ }\textbf {\bibinfo {volume} {C77}},\ \bibinfo {pages} {688}
  (\bibinfo {year} {2017})},\ \Eprint {http://arxiv.org/abs/1704.05447}
  {arXiv:1704.05447 [hep-ph]} \BibitemShut {NoStop}%
\bibitem [{\citenamefont {Hiller}\ and\ \citenamefont
  {Nisandzic}(2017)}]{Hiller:2017bzc}%
  \BibitemOpen
  \bibfield  {author} {\bibinfo {author} {\bibfnamefont {G.}~\bibnamefont
  {Hiller}}\ and\ \bibinfo {author} {\bibfnamefont {I.}~\bibnamefont
  {Nisandzic}},\ }\bibfield  {title} {\enquote {\bibinfo {title} {\emph{$R_K$
  and $R_{K^{*}}$ Beyond the Standard Model}},}\ }\href {\doibase
  10.1103/PhysRevD.96.035003} {\bibfield  {journal} {\bibinfo  {journal} {Phys.
  Rev.}\ }\textbf {\bibinfo {volume} {D96}},\ \bibinfo {pages} {035003}
  (\bibinfo {year} {2017})},\ \Eprint {http://arxiv.org/abs/1704.05444}
  {arXiv:1704.05444 [hep-ph]} \BibitemShut {NoStop}%
\bibitem [{\citenamefont {D'Amico}\ \emph {et~al.}(2017)\citenamefont
  {D'Amico}, \citenamefont {Nardecchia}, \citenamefont {Panci}, \citenamefont
  {Sannino}, \citenamefont {Strumia}, \citenamefont {Torre},\ and\
  \citenamefont {Urbano}}]{DAmico:2017mtc}%
  \BibitemOpen
  \bibfield  {author} {\bibinfo {author} {\bibfnamefont {G.}~\bibnamefont
  {D'Amico}}, \bibinfo {author} {\bibfnamefont {M.}~\bibnamefont {Nardecchia}},
  \bibinfo {author} {\bibfnamefont {P.}~\bibnamefont {Panci}}, \bibinfo
  {author} {\bibfnamefont {F.}~\bibnamefont {Sannino}}, \bibinfo {author}
  {\bibfnamefont {A.}~\bibnamefont {Strumia}}, \bibinfo {author} {\bibfnamefont
  {R.}~\bibnamefont {Torre}}, \ and\ \bibinfo {author} {\bibfnamefont
  {A.}~\bibnamefont {Urbano}},\ }\bibfield  {title} {\enquote {\bibinfo {title}
  {\emph{Flavour Anomalies After the $R_{K^*}$ Measurement}},}\ }\href
  {\doibase 10.1007/JHEP09(2017)010} {\bibfield  {journal} {\bibinfo  {journal}
  {JHEP}\ }\textbf {\bibinfo {volume} {09}},\ \bibinfo {pages} {010} (\bibinfo
  {year} {2017})},\ \Eprint {http://arxiv.org/abs/1704.05438} {arXiv:1704.05438
  [hep-ph]} \BibitemShut {NoStop}%
\bibitem [{\citenamefont {Altmannshofer}\ \emph {et~al.}(2017)\citenamefont
  {Altmannshofer}, \citenamefont {Stangl},\ and\ \citenamefont
  {Straub}}]{Altmannshofer:2017yso}%
  \BibitemOpen
  \bibfield  {author} {\bibinfo {author} {\bibfnamefont {W.}~\bibnamefont
  {Altmannshofer}}, \bibinfo {author} {\bibfnamefont {P.}~\bibnamefont
  {Stangl}}, \ and\ \bibinfo {author} {\bibfnamefont {D.~M.}\ \bibnamefont
  {Straub}},\ }\bibfield  {title} {\enquote {\bibinfo {title}
  {\emph{Interpreting Hints for Lepton Flavor Universality Violation}},}\
  }\href {\doibase 10.1103/PhysRevD.96.055008} {\bibfield  {journal} {\bibinfo
  {journal} {Phys. Rev.}\ }\textbf {\bibinfo {volume} {D96}},\ \bibinfo {pages}
  {055008} (\bibinfo {year} {2017})},\ \Eprint
  {http://arxiv.org/abs/1704.05435} {arXiv:1704.05435 [hep-ph]} \BibitemShut
  {NoStop}%
\bibitem [{\citenamefont {Capdevila}\ \emph {et~al.}(2017)\citenamefont
  {Capdevila}, \citenamefont {Crivellin}, \citenamefont {Descotes-Genon},
  \citenamefont {Matias},\ and\ \citenamefont {Virto}}]{Capdevila:2017bsm}%
  \BibitemOpen
  \bibfield  {author} {\bibinfo {author} {\bibfnamefont {B.}~\bibnamefont
  {Capdevila}}, \bibinfo {author} {\bibfnamefont {A.}~\bibnamefont
  {Crivellin}}, \bibinfo {author} {\bibfnamefont {S.}~\bibnamefont
  {Descotes-Genon}}, \bibinfo {author} {\bibfnamefont {J.}~\bibnamefont
  {Matias}}, \ and\ \bibinfo {author} {\bibfnamefont {J.}~\bibnamefont
  {Virto}},\ }\bibfield  {title} {\enquote {\bibinfo {title} {\emph{Patterns of
  New Physics in $b\to s\ell^+\ell^-$ Transitions in the Light of Recent
  Data}},}\ }\href@noop {} {\  (\bibinfo {year} {2017})},\ \Eprint
  {http://arxiv.org/abs/1704.05340} {arXiv:1704.05340 [hep-ph]} \BibitemShut
  {NoStop}%
\bibitem [{\citenamefont {Becirevic}\ \emph {et~al.}(2016)\citenamefont
  {Becirevic}, \citenamefont {Fajfer}, \citenamefont {Kosnik},\ and\
  \citenamefont {Sumensari}}]{Becirevic:2016yqi}%
  \BibitemOpen
  \bibfield  {author} {\bibinfo {author} {\bibfnamefont {D.}~\bibnamefont
  {Becirevic}}, \bibinfo {author} {\bibfnamefont {S.}~\bibnamefont {Fajfer}},
  \bibinfo {author} {\bibfnamefont {N.}~\bibnamefont {Kosnik}}, \ and\ \bibinfo
  {author} {\bibfnamefont {O.}~\bibnamefont {Sumensari}},\ }\bibfield  {title}
  {\enquote {\bibinfo {title} {\emph{Leptoquark Model to Explain the
  $B$-Physics Anomalies, $R_K$ and $R_D$}},}\ }\href {\doibase
  10.1103/PhysRevD.94.115021} {\bibfield  {journal} {\bibinfo  {journal} {Phys.
  Rev.}\ }\textbf {\bibinfo {volume} {D94}},\ \bibinfo {pages} {115021}
  (\bibinfo {year} {2016})},\ \Eprint {http://arxiv.org/abs/1608.08501}
  {arXiv:1608.08501 [hep-ph]} \BibitemShut {NoStop}%
\bibitem [{\citenamefont {Buttazzo}\ \emph {et~al.}(2017)\citenamefont
  {Buttazzo}, \citenamefont {Greljo}, \citenamefont {Isidori},\ and\
  \citenamefont {Marzocca}}]{Buttazzo:2017ixm}%
  \BibitemOpen
  \bibfield  {author} {\bibinfo {author} {\bibfnamefont {D.}~\bibnamefont
  {Buttazzo}}, \bibinfo {author} {\bibfnamefont {A.}~\bibnamefont {Greljo}},
  \bibinfo {author} {\bibfnamefont {G.}~\bibnamefont {Isidori}}, \ and\
  \bibinfo {author} {\bibfnamefont {D.}~\bibnamefont {Marzocca}},\ }\bibfield
  {title} {\enquote {\bibinfo {title} {\emph{B-Physics Anomalies: A Guide to
  Combined Explanations}},}\ }\href {\doibase 10.1007/JHEP11(2017)044}
  {\bibfield  {journal} {\bibinfo  {journal} {JHEP}\ }\textbf {\bibinfo
  {volume} {11}},\ \bibinfo {pages} {044} (\bibinfo {year} {2017})},\ \Eprint
  {http://arxiv.org/abs/1706.07808} {arXiv:1706.07808 [hep-ph]} \BibitemShut
  {NoStop}%
\bibitem [{\citenamefont {Fajfer}\ and\ \citenamefont
  {Kosnik}(2016)}]{Fajfer:2015ycq}%
  \BibitemOpen
  \bibfield  {author} {\bibinfo {author} {\bibfnamefont {S.}~\bibnamefont
  {Fajfer}}\ and\ \bibinfo {author} {\bibfnamefont {N.}~\bibnamefont
  {Kosnik}},\ }\bibfield  {title} {\enquote {\bibinfo {title} {\emph{Vector
  Leptoquark Resolution of $R_K$ and $R_{D^{(*)}}$ Puzzles}},}\ }\href
  {\doibase 10.1016/j.physletb.2016.02.018} {\bibfield  {journal} {\bibinfo
  {journal} {Phys. Lett.}\ }\textbf {\bibinfo {volume} {B755}},\ \bibinfo
  {pages} {270--274} (\bibinfo {year} {2016})},\ \Eprint
  {http://arxiv.org/abs/1511.06024} {arXiv:1511.06024 [hep-ph]} \BibitemShut
  {NoStop}%
\bibitem [{\citenamefont {Hewett}\ and\ \citenamefont
  {Rizzo}(1989)}]{Hewett:1988xc}%
  \BibitemOpen
  \bibfield  {author} {\bibinfo {author} {\bibfnamefont {J.~L.}\ \bibnamefont
  {Hewett}}\ and\ \bibinfo {author} {\bibfnamefont {T.~G.}\ \bibnamefont
  {Rizzo}},\ }\bibfield  {title} {\enquote {\bibinfo {title} {\emph{Low-Energy
  Phenomenology of Superstring Inspired E(6) Models}},}\ }\href {\doibase
  10.1016/0370-1573(89)90071-9} {\bibfield  {journal} {\bibinfo  {journal}
  {Phys. Rept.}\ }\textbf {\bibinfo {volume} {183}},\ \bibinfo {pages} {193}
  (\bibinfo {year} {1989})}\BibitemShut {NoStop}%
\bibitem [{\citenamefont {Tanaka}\ and\ \citenamefont
  {Watanabe}(1992)}]{Tanaka:1991nr}%
  \BibitemOpen
  \bibfield  {author} {\bibinfo {author} {\bibfnamefont {H.}~\bibnamefont
  {Tanaka}}\ and\ \bibinfo {author} {\bibfnamefont {I.}~\bibnamefont
  {Watanabe}},\ }\bibfield  {title} {\enquote {\bibinfo {title} {\emph{Color
  Sextet Quark Productions at Hadron Colliders}},}\ }\href {\doibase
  10.1142/S0217751X92001204} {\bibfield  {journal} {\bibinfo  {journal} {Int.
  J. Mod. Phys.}\ }\textbf {\bibinfo {volume} {A7}},\ \bibinfo {pages}
  {2679--2694} (\bibinfo {year} {1992})}\BibitemShut {NoStop}%
\bibitem [{\citenamefont {Atag}\ \emph {et~al.}(1999)\citenamefont {Atag},
  \citenamefont {Cakir},\ and\ \citenamefont {Sultansoy}}]{Atag:1998xq}%
  \BibitemOpen
  \bibfield  {author} {\bibinfo {author} {\bibfnamefont {S.}~\bibnamefont
  {Atag}}, \bibinfo {author} {\bibfnamefont {O.}~\bibnamefont {Cakir}}, \ and\
  \bibinfo {author} {\bibfnamefont {S.}~\bibnamefont {Sultansoy}},\ }\bibfield
  {title} {\enquote {\bibinfo {title} {\emph{Resonance Production of Diquarks
  at the CERN LHC}},}\ }\href {\doibase 10.1103/PhysRevD.59.015008} {\bibfield
  {journal} {\bibinfo  {journal} {Phys. Rev.}\ }\textbf {\bibinfo {volume}
  {D59}},\ \bibinfo {pages} {015008} (\bibinfo {year} {1999})}\BibitemShut
  {NoStop}%
\bibitem [{\citenamefont {Cakir}\ and\ \citenamefont
  {Sahin}(2005)}]{Cakir:2005iw}%
  \BibitemOpen
  \bibfield  {author} {\bibinfo {author} {\bibfnamefont {O.}~\bibnamefont
  {Cakir}}\ and\ \bibinfo {author} {\bibfnamefont {M.}~\bibnamefont {Sahin}},\
  }\bibfield  {title} {\enquote {\bibinfo {title} {\emph{Resonant Production of
  Diquarks at High Energy $p p$, ep and $e^{+} e^{-}$ Colliders}},}\ }\href
  {\doibase 10.1103/PhysRevD.72.115011} {\bibfield  {journal} {\bibinfo
  {journal} {Phys. Rev.}\ }\textbf {\bibinfo {volume} {D72}},\ \bibinfo {pages}
  {115011} (\bibinfo {year} {2005})},\ \Eprint
  {http://arxiv.org/abs/hep-ph/0508205} {arXiv:hep-ph/0508205 [hep-ph]}
  \BibitemShut {NoStop}%
\bibitem [{\citenamefont {Chen}\ \emph {et~al.}(2009)\citenamefont {Chen},
  \citenamefont {Klemm}, \citenamefont {Rentala},\ and\ \citenamefont
  {Wang}}]{Chen:2008hh}%
  \BibitemOpen
  \bibfield  {author} {\bibinfo {author} {\bibfnamefont {C.}~\bibnamefont
  {Chen}}, \bibinfo {author} {\bibfnamefont {W.}~\bibnamefont {Klemm}},
  \bibinfo {author} {\bibfnamefont {V.}~\bibnamefont {Rentala}}, \ and\
  \bibinfo {author} {\bibfnamefont {K.}~\bibnamefont {Wang}},\ }\bibfield
  {title} {\enquote {\bibinfo {title} {\emph{Color Sextet Scalars at the CERN
  Large Hadron Collider}},}\ }\href {\doibase 10.1103/PhysRevD.79.054002}
  {\bibfield  {journal} {\bibinfo  {journal} {Phys. Rev.}\ }\textbf {\bibinfo
  {volume} {D79}},\ \bibinfo {pages} {054002} (\bibinfo {year} {2009})},\
  \Eprint {http://arxiv.org/abs/0811.2105} {arXiv:0811.2105 [hep-ph]}
  \BibitemShut {NoStop}%
\bibitem [{\citenamefont {Han}\ \emph {et~al.}(2010)\citenamefont {Han},
  \citenamefont {Lewis},\ and\ \citenamefont {McElmurry}}]{Han:2009ya}%
  \BibitemOpen
  \bibfield  {author} {\bibinfo {author} {\bibfnamefont {T.}~\bibnamefont
  {Han}}, \bibinfo {author} {\bibfnamefont {I.}~\bibnamefont {Lewis}}, \ and\
  \bibinfo {author} {\bibfnamefont {T.}~\bibnamefont {McElmurry}},\ }\bibfield
  {title} {\enquote {\bibinfo {title} {\emph{QCD Corrections to Scalar Diquark
  Production at Hadron Colliders}},}\ }\href {\doibase 10.1007/JHEP01(2010)123}
  {\bibfield  {journal} {\bibinfo  {journal} {JHEP}\ }\textbf {\bibinfo
  {volume} {01}},\ \bibinfo {pages} {123} (\bibinfo {year} {2010})},\ \Eprint
  {http://arxiv.org/abs/0909.2666} {arXiv:0909.2666 [hep-ph]} \BibitemShut
  {NoStop}%
\bibitem [{\citenamefont {Gogoladze}\ \emph {et~al.}(2010)\citenamefont
  {Gogoladze}, \citenamefont {Mimura}, \citenamefont {Okada},\ and\
  \citenamefont {Shafi}}]{Gogoladze:2010xd}%
  \BibitemOpen
  \bibfield  {author} {\bibinfo {author} {\bibfnamefont {I.}~\bibnamefont
  {Gogoladze}}, \bibinfo {author} {\bibfnamefont {Y.}~\bibnamefont {Mimura}},
  \bibinfo {author} {\bibfnamefont {N.}~\bibnamefont {Okada}}, \ and\ \bibinfo
  {author} {\bibfnamefont {Q.}~\bibnamefont {Shafi}},\ }\bibfield  {title}
  {\enquote {\bibinfo {title} {\emph{Color Triplet Diquarks at the LHC}},}\
  }\href {\doibase 10.1016/j.physletb.2010.02.068} {\bibfield  {journal}
  {\bibinfo  {journal} {Phys. Lett.}\ }\textbf {\bibinfo {volume} {B686}},\
  \bibinfo {pages} {233--238} (\bibinfo {year} {2010})},\ \Eprint
  {http://arxiv.org/abs/1001.5260} {arXiv:1001.5260 [hep-ph]} \BibitemShut
  {NoStop}%
\bibitem [{\citenamefont {Berger}\ \emph {et~al.}(2010)\citenamefont {Berger},
  \citenamefont {Cao}, \citenamefont {Chen}, \citenamefont {Shaughnessy},\ and\
  \citenamefont {Zhang}}]{Berger:2010fy}%
  \BibitemOpen
  \bibfield  {author} {\bibinfo {author} {\bibfnamefont {E.~L.}\ \bibnamefont
  {Berger}}, \bibinfo {author} {\bibfnamefont {Q.}~\bibnamefont {Cao}},
  \bibinfo {author} {\bibfnamefont {C.}~\bibnamefont {Chen}}, \bibinfo {author}
  {\bibfnamefont {G.}~\bibnamefont {Shaughnessy}}, \ and\ \bibinfo {author}
  {\bibfnamefont {H.}~\bibnamefont {Zhang}},\ }\bibfield  {title} {\enquote
  {\bibinfo {title} {\emph{Color Sextet Scalars in Early LHC Experiments}},}\
  }\href {\doibase 10.1103/PhysRevLett.105.181802} {\bibfield  {journal}
  {\bibinfo  {journal} {Phys. Rev. Lett.}\ }\textbf {\bibinfo {volume} {105}},\
  \bibinfo {pages} {181802} (\bibinfo {year} {2010})},\ \Eprint
  {http://arxiv.org/abs/1005.2622} {arXiv:1005.2622 [hep-ph]} \BibitemShut
  {NoStop}%
\bibitem [{\citenamefont {Baldes}\ \emph {et~al.}(2011)\citenamefont {Baldes},
  \citenamefont {Bell},\ and\ \citenamefont {Volkas}}]{Baldes:2011mh}%
  \BibitemOpen
  \bibfield  {author} {\bibinfo {author} {\bibfnamefont {I.}~\bibnamefont
  {Baldes}}, \bibinfo {author} {\bibfnamefont {N.~F.}\ \bibnamefont {Bell}}, \
  and\ \bibinfo {author} {\bibfnamefont {R.~R.}\ \bibnamefont {Volkas}},\
  }\bibfield  {title} {\enquote {\bibinfo {title} {\emph{Baryon Number
  Violating Scalar Diquarks at the LHC}},}\ }\href {\doibase
  10.1103/PhysRevD.84.115019} {\bibfield  {journal} {\bibinfo  {journal} {Phys.
  Rev.}\ }\textbf {\bibinfo {volume} {D84}},\ \bibinfo {pages} {115019}
  (\bibinfo {year} {2011})},\ \Eprint {http://arxiv.org/abs/1110.4450}
  {arXiv:1110.4450 [hep-ph]} \BibitemShut {NoStop}%
\bibitem [{\citenamefont {Richardson}\ and\ \citenamefont
  {Winn}(2012)}]{Richardson:2011df}%
  \BibitemOpen
  \bibfield  {author} {\bibinfo {author} {\bibfnamefont {P.}~\bibnamefont
  {Richardson}}\ and\ \bibinfo {author} {\bibfnamefont {D.}~\bibnamefont
  {Winn}},\ }\bibfield  {title} {\enquote {\bibinfo {title} {\emph{Simulation
  of Sextet Diquark Production}},}\ }\href {\doibase
  10.1140/epjc/s10052-012-1862-z} {\bibfield  {journal} {\bibinfo  {journal}
  {Eur. Phys. J.}\ }\textbf {\bibinfo {volume} {C72}},\ \bibinfo {pages} {1862}
  (\bibinfo {year} {2012})},\ \Eprint {http://arxiv.org/abs/1108.6154}
  {arXiv:1108.6154 [hep-ph]} \BibitemShut {NoStop}%
\bibitem [{\citenamefont {Karabacak}\ \emph {et~al.}(2012)\citenamefont
  {Karabacak}, \citenamefont {Nandi},\ and\ \citenamefont
  {Rai}}]{Karabacak:2012rn}%
  \BibitemOpen
  \bibfield  {author} {\bibinfo {author} {\bibfnamefont {D.}~\bibnamefont
  {Karabacak}}, \bibinfo {author} {\bibfnamefont {S.}~\bibnamefont {Nandi}}, \
  and\ \bibinfo {author} {\bibfnamefont {S.~K.}\ \bibnamefont {Rai}},\
  }\bibfield  {title} {\enquote {\bibinfo {title} {\emph{Diquark Resonance and
  Single Top Production at the Large Hadron Collider}},}\ }\href {\doibase
  10.1103/PhysRevD.85.075011} {\bibfield  {journal} {\bibinfo  {journal} {Phys.
  Rev.}\ }\textbf {\bibinfo {volume} {D85}},\ \bibinfo {pages} {075011}
  (\bibinfo {year} {2012})},\ \Eprint {http://arxiv.org/abs/1201.2917}
  {arXiv:1201.2917 [hep-ph]} \BibitemShut {NoStop}%
\bibitem [{\citenamefont {Kohda}\ \emph {et~al.}(2013)\citenamefont {Kohda},
  \citenamefont {Sugiyama},\ and\ \citenamefont {Tsumura}}]{Kohda:2012sr}%
  \BibitemOpen
  \bibfield  {author} {\bibinfo {author} {\bibfnamefont {M.}~\bibnamefont
  {Kohda}}, \bibinfo {author} {\bibfnamefont {H.}~\bibnamefont {Sugiyama}}, \
  and\ \bibinfo {author} {\bibfnamefont {K.}~\bibnamefont {Tsumura}},\
  }\bibfield  {title} {\enquote {\bibinfo {title} {\emph{Lepton Number
  Violation at the LHC with Leptoquark and Diquark}},}\ }\href {\doibase
  10.1016/j.physletb.2012.12.048} {\bibfield  {journal} {\bibinfo  {journal}
  {Phys. Lett.}\ }\textbf {\bibinfo {volume} {B718}},\ \bibinfo {pages}
  {1436--1440} (\bibinfo {year} {2013})},\ \Eprint
  {http://arxiv.org/abs/1210.5622} {arXiv:1210.5622 [hep-ph]} \BibitemShut
  {NoStop}%
\bibitem [{\citenamefont {Das}\ \emph {et~al.}(2015)\citenamefont {Das},
  \citenamefont {Majhi}, \citenamefont {Rai},\ and\ \citenamefont
  {Shivaji}}]{Das:2015lna}%
  \BibitemOpen
  \bibfield  {author} {\bibinfo {author} {\bibfnamefont {K.}~\bibnamefont
  {Das}}, \bibinfo {author} {\bibfnamefont {S.}~\bibnamefont {Majhi}}, \bibinfo
  {author} {\bibfnamefont {Santosh~K.}\ \bibnamefont {Rai}}, \ and\ \bibinfo
  {author} {\bibfnamefont {A.}~\bibnamefont {Shivaji}},\ }\bibfield  {title}
  {\enquote {\bibinfo {title} {\emph{NLO QCD Corrections to the Resonant Vector
  Diquark Production at the LHC}},}\ }\href {\doibase 10.1007/JHEP10(2015)122}
  {\bibfield  {journal} {\bibinfo  {journal} {JHEP}\ }\textbf {\bibinfo
  {volume} {10}},\ \bibinfo {pages} {122} (\bibinfo {year} {2015})},\ \Eprint
  {http://arxiv.org/abs/1505.07256} {arXiv:1505.07256 [hep-ph]} \BibitemShut
  {NoStop}%
\bibitem [{\citenamefont {Chivukula}\ \emph {et~al.}(2015)\citenamefont
  {Chivukula}, \citenamefont {Ittisamai}, \citenamefont {Mohan},\ and\
  \citenamefont {Simmons}}]{Chivukula:2015zma}%
  \BibitemOpen
  \bibfield  {author} {\bibinfo {author} {\bibfnamefont {R.~S.}\ \bibnamefont
  {Chivukula}}, \bibinfo {author} {\bibfnamefont {P.}~\bibnamefont
  {Ittisamai}}, \bibinfo {author} {\bibfnamefont {K.}~\bibnamefont {Mohan}}, \
  and\ \bibinfo {author} {\bibfnamefont {E.~H.}\ \bibnamefont {Simmons}},\
  }\bibfield  {title} {\enquote {\bibinfo {title} {\emph{Color Discriminant
  Variable and Scalar Diquarks at the LHC}},}\ }\href {\doibase
  10.1103/PhysRevD.92.075020} {\bibfield  {journal} {\bibinfo  {journal} {Phys.
  Rev.}\ }\textbf {\bibinfo {volume} {D92}},\ \bibinfo {pages} {075020}
  (\bibinfo {year} {2015})},\ \Eprint {http://arxiv.org/abs/1507.06676}
  {arXiv:1507.06676 [hep-ph]} \BibitemShut {NoStop}%
\bibitem [{\citenamefont {Zhan}\ \emph {et~al.}(2014)\citenamefont {Zhan},
  \citenamefont {Liu}, \citenamefont {Li}, \citenamefont {Li},\ and\
  \citenamefont {Li}}]{Zhan:2013sza}%
  \BibitemOpen
  \bibfield  {author} {\bibinfo {author} {\bibfnamefont {Y.}~\bibnamefont
  {Zhan}}, \bibinfo {author} {\bibfnamefont {Z.~L.}\ \bibnamefont {Liu}},
  \bibinfo {author} {\bibfnamefont {S.~A.}\ \bibnamefont {Li}}, \bibinfo
  {author} {\bibfnamefont {C.~S.}\ \bibnamefont {Li}}, \ and\ \bibinfo {author}
  {\bibfnamefont {H.~T.}\ \bibnamefont {Li}},\ }\bibfield  {title} {\enquote
  {\bibinfo {title} {\emph{Threshold Resummation for the Production of a Color
  Sextet (Antitriplet) Scalar at the LHC}},}\ }\href {\doibase
  10.1140/epjc/s10052-014-2716-7} {\bibfield  {journal} {\bibinfo  {journal}
  {Eur. Phys. J.}\ }\textbf {\bibinfo {volume} {C74}},\ \bibinfo {pages} {2716}
  (\bibinfo {year} {2014})},\ \Eprint {http://arxiv.org/abs/1305.5152}
  {arXiv:1305.5152 [hep-ph]} \BibitemShut {NoStop}%
\bibitem [{\citenamefont {Mohapatra}\ and\ \citenamefont
  {Marshak}(1980)}]{Mohapatra:1980qe}%
  \BibitemOpen
  \bibfield  {author} {\bibinfo {author} {\bibfnamefont {R.~N.}\ \bibnamefont
  {Mohapatra}}\ and\ \bibinfo {author} {\bibfnamefont {R.~E.}\ \bibnamefont
  {Marshak}},\ }\bibfield  {title} {\enquote {\bibinfo {title} {\emph{Local B-L
  Symmetry of Electroweak Interactions, Majorana Neutrinos and Neutron
  Oscillations}},}\ }\href {\doibase 10.1103/PhysRevLett.44.1316} {\bibfield
  {journal} {\bibinfo  {journal} {Phys. Rev. Lett.}\ }\textbf {\bibinfo
  {volume} {44}},\ \bibinfo {pages} {1316--1319} (\bibinfo {year} {1980})},\
  \bibinfo {note} {[Erratum: Phys. Rev. Lett.44,1643(1980)]}\BibitemShut
  {NoStop}%
\bibitem [{\citenamefont {Babu}\ \emph {et~al.}(2009)\citenamefont {Babu},
  \citenamefont {Dev},\ and\ \citenamefont {Mohapatra}}]{Babu:2008rq}%
  \BibitemOpen
  \bibfield  {author} {\bibinfo {author} {\bibfnamefont {K.~S.}\ \bibnamefont
  {Babu}}, \bibinfo {author} {\bibfnamefont {P.~S.~B.}\ \bibnamefont {Dev}}, \
  and\ \bibinfo {author} {\bibfnamefont {R.~N.}\ \bibnamefont {Mohapatra}},\
  }\bibfield  {title} {\enquote {\bibinfo {title} {\emph{Neutrino Mass
  Hierarchy, Neutron-Antineutron Oscillation from Baryogenesis}},}\ }\href
  {\doibase 10.1103/PhysRevD.79.015017} {\bibfield  {journal} {\bibinfo
  {journal} {Phys. Rev.}\ }\textbf {\bibinfo {volume} {D79}},\ \bibinfo {pages}
  {015017} (\bibinfo {year} {2009})},\ \Eprint {http://arxiv.org/abs/0811.3411}
  {arXiv:0811.3411 [hep-ph]} \BibitemShut {NoStop}%
\bibitem [{\citenamefont {Ajaib}\ \emph {et~al.}(2009)\citenamefont {Ajaib},
  \citenamefont {Gogoladze}, \citenamefont {Mimura},\ and\ \citenamefont
  {Shafi}}]{Ajaib:2009fq}%
  \BibitemOpen
  \bibfield  {author} {\bibinfo {author} {\bibfnamefont {M.~A.}\ \bibnamefont
  {Ajaib}}, \bibinfo {author} {\bibfnamefont {I.}~\bibnamefont {Gogoladze}},
  \bibinfo {author} {\bibfnamefont {Y.}~\bibnamefont {Mimura}}, \ and\ \bibinfo
  {author} {\bibfnamefont {Q.}~\bibnamefont {Shafi}},\ }\bibfield  {title}
  {\enquote {\bibinfo {title} {\emph{Observable $n\bar{n}$ Oscillations with
  New Physics at LHC}},}\ }\href {\doibase 10.1103/PhysRevD.80.125026}
  {\bibfield  {journal} {\bibinfo  {journal} {Phys. Rev.}\ }\textbf {\bibinfo
  {volume} {D80}},\ \bibinfo {pages} {125026} (\bibinfo {year} {2009})},\
  \Eprint {http://arxiv.org/abs/0910.1877} {arXiv:0910.1877 [hep-ph]}
  \BibitemShut {NoStop}%
\bibitem [{\citenamefont {Gu}\ and\ \citenamefont {Sarkar}(2011)}]{Gu:2011ff}%
  \BibitemOpen
  \bibfield  {author} {\bibinfo {author} {\bibfnamefont {P.-H.}\ \bibnamefont
  {Gu}}\ and\ \bibinfo {author} {\bibfnamefont {U.}~\bibnamefont {Sarkar}},\
  }\bibfield  {title} {\enquote {\bibinfo {title} {\emph{Baryogenesis and
  Neutron-Antineutron Oscillation at TeV}},}\ }\href {\doibase
  10.1016/j.physletb.2011.10.017} {\bibfield  {journal} {\bibinfo  {journal}
  {Phys. Lett.}\ }\textbf {\bibinfo {volume} {B705}},\ \bibinfo {pages}
  {170--173} (\bibinfo {year} {2011})},\ \Eprint
  {http://arxiv.org/abs/1107.0173} {arXiv:1107.0173 [hep-ph]} \BibitemShut
  {NoStop}%
\bibitem [{\citenamefont {Babu}\ and\ \citenamefont
  {Mohapatra}(2012)}]{Babu:2012vc}%
  \BibitemOpen
  \bibfield  {author} {\bibinfo {author} {\bibfnamefont {K.~S.}\ \bibnamefont
  {Babu}}\ and\ \bibinfo {author} {\bibfnamefont {R.~N.}\ \bibnamefont
  {Mohapatra}},\ }\bibfield  {title} {\enquote {\bibinfo {title}
  {\emph{Coupling Unification, GUT-Scale Baryogenesis and Neutron-Antineutron
  Oscillation in SO(10)}},}\ }\href {\doibase 10.1016/j.physletb.2012.08.006}
  {\bibfield  {journal} {\bibinfo  {journal} {Phys. Lett.}\ }\textbf {\bibinfo
  {volume} {B715}},\ \bibinfo {pages} {328--334} (\bibinfo {year} {2012})},\
  \Eprint {http://arxiv.org/abs/1206.5701} {arXiv:1206.5701 [hep-ph]}
  \BibitemShut {NoStop}%
\bibitem [{\citenamefont {Babu}\ \emph {et~al.}(2013)\citenamefont {Babu},
  \citenamefont {Dev}, \citenamefont {Fortes},\ and\ \citenamefont
  {Mohapatra}}]{Babu:2013yca}%
  \BibitemOpen
  \bibfield  {author} {\bibinfo {author} {\bibfnamefont {K.~S.}\ \bibnamefont
  {Babu}}, \bibinfo {author} {\bibfnamefont {P.~S.~B.}\ \bibnamefont {Dev}},
  \bibinfo {author} {\bibfnamefont {E.~C. F.~S.}\ \bibnamefont {Fortes}}, \
  and\ \bibinfo {author} {\bibfnamefont {R.~N.}\ \bibnamefont {Mohapatra}},\
  }\bibfield  {title} {\enquote {\bibinfo {title} {\emph{Post-Sphaleron
  Baryogenesis and an Upper Limit on the Neutron-Antineutron Oscillation
  Time}},}\ }\href {\doibase 10.1103/PhysRevD.87.115019} {\bibfield  {journal}
  {\bibinfo  {journal} {Phys. Rev.}\ }\textbf {\bibinfo {volume} {D87}},\
  \bibinfo {pages} {115019} (\bibinfo {year} {2013})},\ \Eprint
  {http://arxiv.org/abs/1303.6918} {arXiv:1303.6918 [hep-ph]} \BibitemShut
  {NoStop}%
\bibitem [{\citenamefont {Arik}\ \emph {et~al.}(2002)\citenamefont {Arik},
  \citenamefont {Cakir}, \citenamefont {Cetin},\ and\ \citenamefont
  {Sultansoy}}]{Arik:2001bc}%
  \BibitemOpen
  \bibfield  {author} {\bibinfo {author} {\bibfnamefont {E.}~\bibnamefont
  {Arik}}, \bibinfo {author} {\bibfnamefont {O.}~\bibnamefont {Cakir}},
  \bibinfo {author} {\bibfnamefont {S.~A.}\ \bibnamefont {Cetin}}, \ and\
  \bibinfo {author} {\bibfnamefont {S.}~\bibnamefont {Sultansoy}},\ }\bibfield
  {title} {\enquote {\bibinfo {title} {\emph{A Search for Vector Diquarks at
  the CERN LHC}},}\ }\href {\doibase 10.1088/1126-6708/2002/09/024} {\bibfield
  {journal} {\bibinfo  {journal} {JHEP}\ }\textbf {\bibinfo {volume} {09}},\
  \bibinfo {pages} {024} (\bibinfo {year} {2002})},\ \Eprint
  {http://arxiv.org/abs/hep-ph/0109011} {arXiv:hep-ph/0109011 [hep-ph]}
  \BibitemShut {NoStop}%
\bibitem [{\citenamefont {Sahin}\ and\ \citenamefont
  {Cakir}(2009)}]{Sahin:2009dca}%
  \BibitemOpen
  \bibfield  {author} {\bibinfo {author} {\bibfnamefont {M.}~\bibnamefont
  {Sahin}}\ and\ \bibinfo {author} {\bibfnamefont {O.}~\bibnamefont {Cakir}},\
  }\bibfield  {title} {\enquote {\bibinfo {title} {\emph{Search for Scalar and
  Vector Diquarks at the LHC}},}\ }\href@noop {} {\bibfield  {journal}
  {\bibinfo  {journal} {Balk. Phys. Lett.}\ }\textbf {\bibinfo {volume} {17}},\
  \bibinfo {pages} {132--137} (\bibinfo {year} {2009})}\BibitemShut {NoStop}%
\bibitem [{\citenamefont {Zhang}\ \emph {et~al.}(2011)\citenamefont {Zhang},
  \citenamefont {Berger}, \citenamefont {Cao}, \citenamefont {Chen},\ and\
  \citenamefont {Shaughnessy}}]{Zhang:2010kr}%
  \BibitemOpen
  \bibfield  {author} {\bibinfo {author} {\bibfnamefont {H.}~\bibnamefont
  {Zhang}}, \bibinfo {author} {\bibfnamefont {E.~L.}\ \bibnamefont {Berger}},
  \bibinfo {author} {\bibfnamefont {Q.-H.}\ \bibnamefont {Cao}}, \bibinfo
  {author} {\bibfnamefont {C.-R.}\ \bibnamefont {Chen}}, \ and\ \bibinfo
  {author} {\bibfnamefont {G.}~\bibnamefont {Shaughnessy}},\ }\bibfield
  {title} {\enquote {\bibinfo {title} {\emph{Color Sextet Vector Bosons and
  Same-Sign Top Quark Pairs at the LHC}},}\ }\href {\doibase
  10.1016/j.physletb.2010.12.005} {\bibfield  {journal} {\bibinfo  {journal}
  {Phys. Lett.}\ }\textbf {\bibinfo {volume} {B696}},\ \bibinfo {pages}
  {68--73} (\bibinfo {year} {2011})},\ \Eprint {http://arxiv.org/abs/1009.5379}
  {arXiv:1009.5379 [hep-ph]} \BibitemShut {NoStop}%
\bibitem [{\citenamefont {Grinstein}\ \emph
  {et~al.}(2011{\natexlab{a}})\citenamefont {Grinstein}, \citenamefont {Kagan},
  \citenamefont {Trott},\ and\ \citenamefont {Zupan}}]{Grinstein:2011yv}%
  \BibitemOpen
  \bibfield  {author} {\bibinfo {author} {\bibfnamefont {B.}~\bibnamefont
  {Grinstein}}, \bibinfo {author} {\bibfnamefont {A.~L.}\ \bibnamefont
  {Kagan}}, \bibinfo {author} {\bibfnamefont {M.}~\bibnamefont {Trott}}, \ and\
  \bibinfo {author} {\bibfnamefont {J.}~\bibnamefont {Zupan}},\ }\bibfield
  {title} {\enquote {\bibinfo {title} {\emph{Forward-Backward Asymmetry in $t
  \bar{t}$ Production From Flavour Symmetries}},}\ }\href {\doibase
  10.1103/PhysRevLett.107.012002} {\bibfield  {journal} {\bibinfo  {journal}
  {Phys. Rev. Lett.}\ }\textbf {\bibinfo {volume} {107}},\ \bibinfo {pages}
  {012002} (\bibinfo {year} {2011}{\natexlab{a}})},\ \Eprint
  {http://arxiv.org/abs/1102.3374} {arXiv:1102.3374 [hep-ph]} \BibitemShut
  {NoStop}%
\bibitem [{\citenamefont {Grinstein}\ \emph
  {et~al.}(2011{\natexlab{b}})\citenamefont {Grinstein}, \citenamefont {Kagan},
  \citenamefont {Zupan},\ and\ \citenamefont {Trott}}]{Grinstein:2011dz}%
  \BibitemOpen
  \bibfield  {author} {\bibinfo {author} {\bibfnamefont {B.}~\bibnamefont
  {Grinstein}}, \bibinfo {author} {\bibfnamefont {A.~L.}\ \bibnamefont
  {Kagan}}, \bibinfo {author} {\bibfnamefont {J.}~\bibnamefont {Zupan}}, \ and\
  \bibinfo {author} {\bibfnamefont {M.}~\bibnamefont {Trott}},\ }\bibfield
  {title} {\enquote {\bibinfo {title} {\emph{Flavor Symmetric Sectors and
  Collider Physics}},}\ }\href {\doibase 10.1007/JHEP10(2011)072} {\bibfield
  {journal} {\bibinfo  {journal} {JHEP}\ }\textbf {\bibinfo {volume} {10}},\
  \bibinfo {pages} {072} (\bibinfo {year} {2011}{\natexlab{b}})},\ \Eprint
  {http://arxiv.org/abs/1108.4027} {arXiv:1108.4027 [hep-ph]} \BibitemShut
  {NoStop}%
\bibitem [{\citenamefont {Gripaios}(2010)}]{Gripaios:2009dq}%
  \BibitemOpen
  \bibfield  {author} {\bibinfo {author} {\bibfnamefont {B.}~\bibnamefont
  {Gripaios}},\ }\bibfield  {title} {\enquote {\bibinfo {title}
  {\emph{Composite Leptoquarks at the LHC}},}\ }\href {\doibase
  10.1007/JHEP02(2010)045} {\bibfield  {journal} {\bibinfo  {journal} {JHEP}\
  }\textbf {\bibinfo {volume} {02}},\ \bibinfo {pages} {045} (\bibinfo {year}
  {2010})},\ \Eprint {http://arxiv.org/abs/0910.1789} {arXiv:0910.1789
  [hep-ph]} \BibitemShut {NoStop}%
\bibitem [{\citenamefont {Gripaios}\ \emph {et~al.}(2015)\citenamefont
  {Gripaios}, \citenamefont {Nardecchia},\ and\ \citenamefont
  {Renner}}]{Gripaios:2014tna}%
  \BibitemOpen
  \bibfield  {author} {\bibinfo {author} {\bibfnamefont {B.}~\bibnamefont
  {Gripaios}}, \bibinfo {author} {\bibfnamefont {M.}~\bibnamefont
  {Nardecchia}}, \ and\ \bibinfo {author} {\bibfnamefont {S.~A.}\ \bibnamefont
  {Renner}},\ }\bibfield  {title} {\enquote {\bibinfo {title} {\emph{Composite
  Leptoquarks and Anomalies in $B$-Meson Decays}},}\ }\href {\doibase
  10.1007/JHEP05(2015)006} {\bibfield  {journal} {\bibinfo  {journal} {JHEP}\
  }\textbf {\bibinfo {volume} {05}},\ \bibinfo {pages} {006} (\bibinfo {year}
  {2015})},\ \Eprint {http://arxiv.org/abs/1412.1791} {arXiv:1412.1791
  [hep-ph]} \BibitemShut {NoStop}%
\bibitem [{\citenamefont {Barbieri}\ \emph {et~al.}(2017)\citenamefont
  {Barbieri}, \citenamefont {Murphy},\ and\ \citenamefont
  {Senia}}]{Barbieri:2016las}%
  \BibitemOpen
  \bibfield  {author} {\bibinfo {author} {\bibfnamefont {R.}~\bibnamefont
  {Barbieri}}, \bibinfo {author} {\bibfnamefont {C.~W.}\ \bibnamefont
  {Murphy}}, \ and\ \bibinfo {author} {\bibfnamefont {F.}~\bibnamefont
  {Senia}},\ }\bibfield  {title} {\enquote {\bibinfo {title} {\emph{B-Decay
  Anomalies in a Composite Leptoquark Model}},}\ }\href {\doibase
  10.1140/epjc/s10052-016-4578-7} {\bibfield  {journal} {\bibinfo  {journal}
  {Eur. Phys. J.}\ }\textbf {\bibinfo {volume} {C77}},\ \bibinfo {pages} {8}
  (\bibinfo {year} {2017})},\ \Eprint {http://arxiv.org/abs/1611.04930}
  {arXiv:1611.04930 [hep-ph]} \BibitemShut {NoStop}%
\bibitem [{\citenamefont {Pati}\ and\ \citenamefont
  {Salam}(1974)}]{Pati:1974yy}%
  \BibitemOpen
  \bibfield  {author} {\bibinfo {author} {\bibfnamefont {J.~C.}\ \bibnamefont
  {Pati}}\ and\ \bibinfo {author} {\bibfnamefont {A.}~\bibnamefont {Salam}},\
  }\bibfield  {title} {\enquote {\bibinfo {title} {\emph{Lepton Number as the
  Fourth Color}},}\ }\href {\doibase 10.1103/PhysRevD.10.275,
  10.1103/PhysRevD.11.703.2} {\bibfield  {journal} {\bibinfo  {journal} {Phys.
  Rev.}\ }\textbf {\bibinfo {volume} {D10}},\ \bibinfo {pages} {275--289}
  (\bibinfo {year} {1974})},\ \bibinfo {note} {[Erratum: Phys. Rev. D11, 703
  (1975)]}\BibitemShut {NoStop}%
\bibitem [{\citenamefont {Fileviez~Perez}\ and\ \citenamefont
  {Wise}(2013)}]{Perez:2013osa}%
  \BibitemOpen
  \bibfield  {author} {\bibinfo {author} {\bibfnamefont {P.}~\bibnamefont
  {Fileviez~Perez}}\ and\ \bibinfo {author} {\bibfnamefont {M.~B.}\
  \bibnamefont {Wise}},\ }\bibfield  {title} {\enquote {\bibinfo {title}
  {\emph{Low Scale Quark-Lepton Unification}},}\ }\href {\doibase
  10.1103/PhysRevD.88.057703} {\bibfield  {journal} {\bibinfo  {journal} {Phys.
  Rev.}\ }\textbf {\bibinfo {volume} {D88}},\ \bibinfo {pages} {057703}
  (\bibinfo {year} {2013})},\ \Eprint {http://arxiv.org/abs/1307.6213}
  {arXiv:1307.6213 [hep-ph]} \BibitemShut {NoStop}%
\bibitem [{\citenamefont {Fornal}\ \emph {et~al.}(2015)\citenamefont {Fornal},
  \citenamefont {Rajaraman},\ and\ \citenamefont {Tait}}]{Fornal:2015boa}%
  \BibitemOpen
  \bibfield  {author} {\bibinfo {author} {\bibfnamefont {B.}~\bibnamefont
  {Fornal}}, \bibinfo {author} {\bibfnamefont {A.}~\bibnamefont {Rajaraman}}, \
  and\ \bibinfo {author} {\bibfnamefont {T.~M.~P.}\ \bibnamefont {Tait}},\
  }\bibfield  {title} {\enquote {\bibinfo {title} {\emph{Baryon Number as the
  Fourth Color}},}\ }\href {\doibase 10.1103/PhysRevD.92.055022} {\bibfield
  {journal} {\bibinfo  {journal} {Phys. Rev.}\ }\textbf {\bibinfo {volume}
  {D92}},\ \bibinfo {pages} {055022} (\bibinfo {year} {2015})},\ \Eprint
  {http://arxiv.org/abs/1506.06131} {arXiv:1506.06131 [hep-ph]} \BibitemShut
  {NoStop}%
\bibitem [{\citenamefont {Grinstein}\ \emph {et~al.}(1989)\citenamefont
  {Grinstein}, \citenamefont {Savage},\ and\ \citenamefont
  {Wise}}]{Grinstein:1988me}%
  \BibitemOpen
  \bibfield  {author} {\bibinfo {author} {\bibfnamefont {B.}~\bibnamefont
  {Grinstein}}, \bibinfo {author} {\bibfnamefont {M.~J.}\ \bibnamefont
  {Savage}}, \ and\ \bibinfo {author} {\bibfnamefont {M.~B.}\ \bibnamefont
  {Wise}},\ }\bibfield  {title} {\enquote {\bibinfo {title} {\emph{$B
  \rightarrow X(s) e^+ e^-$ in the Six Quark Model}},}\ }\href {\doibase
  10.1016/0550-3213(89)90078-3} {\bibfield  {journal} {\bibinfo  {journal}
  {Nucl. Phys.}\ }\textbf {\bibinfo {volume} {B319}},\ \bibinfo {pages}
  {271--290} (\bibinfo {year} {1989})}\BibitemShut {NoStop}%
\bibitem [{\citenamefont {Buchalla}\ \emph {et~al.}(1996)\citenamefont
  {Buchalla}, \citenamefont {Buras},\ and\ \citenamefont
  {Lautenbacher}}]{Buchalla:1995vs}%
  \BibitemOpen
  \bibfield  {author} {\bibinfo {author} {\bibfnamefont {G.}~\bibnamefont
  {Buchalla}}, \bibinfo {author} {\bibfnamefont {A.~J.}\ \bibnamefont {Buras}},
  \ and\ \bibinfo {author} {\bibfnamefont {M.~E.}\ \bibnamefont
  {Lautenbacher}},\ }\bibfield  {title} {\enquote {\bibinfo {title} {\emph{Weak
  Decays Beyond Leading Logarithms}},}\ }\href {\doibase
  10.1103/RevModPhys.68.1125} {\bibfield  {journal} {\bibinfo  {journal} {Rev.
  Mod. Phys.}\ }\textbf {\bibinfo {volume} {68}},\ \bibinfo {pages}
  {1125--1144} (\bibinfo {year} {1996})},\ \Eprint
  {http://arxiv.org/abs/hep-ph/9512380} {arXiv:hep-ph/9512380 [hep-ph]}
  \BibitemShut {NoStop}%
\bibitem [{\citenamefont {Alonso}\ \emph {et~al.}(2014)\citenamefont {Alonso},
  \citenamefont {Grinstein},\ and\ \citenamefont
  {Martin~Camalich}}]{Alonso:2014csa}%
  \BibitemOpen
  \bibfield  {author} {\bibinfo {author} {\bibfnamefont {R.}~\bibnamefont
  {Alonso}}, \bibinfo {author} {\bibfnamefont {B.}~\bibnamefont {Grinstein}}, \
  and\ \bibinfo {author} {\bibfnamefont {J.}~\bibnamefont {Martin~Camalich}},\
  }\bibfield  {title} {\enquote {\bibinfo {title} {\emph{$SU(2)\times U(1)$
  Gauge Invariance and the Shape of New Physics in Rare $B$ Decays}},}\ }\href
  {\doibase 10.1103/PhysRevLett.113.241802} {\bibfield  {journal} {\bibinfo
  {journal} {Phys. Rev. Lett.}\ }\textbf {\bibinfo {volume} {113}},\ \bibinfo
  {pages} {241802} (\bibinfo {year} {2014})},\ \Eprint
  {http://arxiv.org/abs/1407.7044} {arXiv:1407.7044 [hep-ph]} \BibitemShut
  {NoStop}%
\bibitem [{\citenamefont {Buras}\ \emph {et~al.}(2015)\citenamefont {Buras},
  \citenamefont {Girrbach-Noe}, \citenamefont {Niehoff},\ and\ \citenamefont
  {Straub}}]{Buras:2014fpa}%
  \BibitemOpen
  \bibfield  {author} {\bibinfo {author} {\bibfnamefont {A.~J.}\ \bibnamefont
  {Buras}}, \bibinfo {author} {\bibfnamefont {J.}~\bibnamefont {Girrbach-Noe}},
  \bibinfo {author} {\bibfnamefont {C.}~\bibnamefont {Niehoff}}, \ and\
  \bibinfo {author} {\bibfnamefont {D.~M.}\ \bibnamefont {Straub}},\ }\bibfield
   {title} {\enquote {\bibinfo {title} {\emph{$ B\to {K}^{\left(\ast
  \right)}\nu \overline{\nu} $ Decays in the Standard Model and Beyond}},}\
  }\href {\doibase 10.1007/JHEP02(2015)184} {\bibfield  {journal} {\bibinfo
  {journal} {JHEP}\ }\textbf {\bibinfo {volume} {02}},\ \bibinfo {pages} {184}
  (\bibinfo {year} {2015})},\ \Eprint {http://arxiv.org/abs/1409.4557}
  {arXiv:1409.4557 [hep-ph]} \BibitemShut {NoStop}%
\bibitem [{\citenamefont {Calibbi}\ \emph {et~al.}(2015)\citenamefont
  {Calibbi}, \citenamefont {Crivellin},\ and\ \citenamefont
  {Ota}}]{Calibbi:2015kma}%
  \BibitemOpen
  \bibfield  {author} {\bibinfo {author} {\bibfnamefont {L.}~\bibnamefont
  {Calibbi}}, \bibinfo {author} {\bibfnamefont {A.}~\bibnamefont {Crivellin}},
  \ and\ \bibinfo {author} {\bibfnamefont {T.}~\bibnamefont {Ota}},\ }\bibfield
   {title} {\enquote {\bibinfo {title} {\emph{Effective Field Theory Approach
  to $b\to s \ell\ell^{(\prime)}$, $B\to K^{(*)}\nu\overline{\nu}$ and $B\to
  D^{(*)}\tau\nu$ with Third Generation Couplings}},}\ }\href {\doibase
  10.1103/PhysRevLett.115.181801} {\bibfield  {journal} {\bibinfo  {journal}
  {Phys. Rev. Lett.}\ }\textbf {\bibinfo {volume} {115}},\ \bibinfo {pages}
  {181801} (\bibinfo {year} {2015})},\ \Eprint
  {http://arxiv.org/abs/1506.02661} {arXiv:1506.02661 [hep-ph]} \BibitemShut
  {NoStop}%
\bibitem [{\citenamefont {Feruglio}\ \emph {et~al.}(2017)\citenamefont
  {Feruglio}, \citenamefont {Paradisi},\ and\ \citenamefont
  {Pattori}}]{Feruglio:2017rjo}%
  \BibitemOpen
  \bibfield  {author} {\bibinfo {author} {\bibfnamefont {F.}~\bibnamefont
  {Feruglio}}, \bibinfo {author} {\bibfnamefont {P.}~\bibnamefont {Paradisi}},
  \ and\ \bibinfo {author} {\bibfnamefont {A.}~\bibnamefont {Pattori}},\
  }\bibfield  {title} {\enquote {\bibinfo {title} {\emph{On the Importance of
  Electroweak Corrections for $B$ Anomalies}},}\ }\href {\doibase
  10.1007/JHEP09(2017)061} {\bibfield  {journal} {\bibinfo  {journal} {JHEP}\
  }\textbf {\bibinfo {volume} {09}},\ \bibinfo {pages} {061} (\bibinfo {year}
  {2017})},\ \Eprint {http://arxiv.org/abs/1705.00929} {arXiv:1705.00929
  [hep-ph]} \BibitemShut {NoStop}%
\bibitem [{\citenamefont {Lutz}\ \emph {et~al.}(2013)\citenamefont {Lutz} \emph
  {et~al.}}]{Lutz:2013ftz}%
  \BibitemOpen
  \bibfield  {author} {\bibinfo {author} {\bibfnamefont {O.}~\bibnamefont
  {Lutz}} \emph {et~al.} (\bibinfo {collaboration} {Belle}),\ }\bibfield
  {title} {\enquote {\bibinfo {title} {\emph{Search for $B \to h^{(*)} \nu
  \bar{\nu}$ with the Full Belle $\Upsilon(4S)$ Data Sample}},}\ }\href
  {\doibase 10.1103/PhysRevD.87.111103} {\bibfield  {journal} {\bibinfo
  {journal} {Phys. Rev.}\ }\textbf {\bibinfo {volume} {D87}},\ \bibinfo {pages}
  {111103} (\bibinfo {year} {2013})},\ \Eprint {http://arxiv.org/abs/1303.3719}
  {arXiv:1303.3719 [hep-ex]} \BibitemShut {NoStop}%
\bibitem [{\citenamefont {Sakaki}\ \emph {et~al.}(2013)\citenamefont {Sakaki},
  \citenamefont {Tanaka}, \citenamefont {Tayduganov},\ and\ \citenamefont
  {Watanabe}}]{Sakaki:2013bfa}%
  \BibitemOpen
  \bibfield  {author} {\bibinfo {author} {\bibfnamefont {Y.}~\bibnamefont
  {Sakaki}}, \bibinfo {author} {\bibfnamefont {M.}~\bibnamefont {Tanaka}},
  \bibinfo {author} {\bibfnamefont {A.}~\bibnamefont {Tayduganov}}, \ and\
  \bibinfo {author} {\bibfnamefont {R.}~\bibnamefont {Watanabe}},\ }\bibfield
  {title} {\enquote {\bibinfo {title} {\emph{Testing Leptoquark Models in $\bar
  B \to D^{(*)} \tau \bar\nu$}},}\ }\href {\doibase 10.1103/PhysRevD.88.094012}
  {\bibfield  {journal} {\bibinfo  {journal} {Phys. Rev.}\ }\textbf {\bibinfo
  {volume} {D88}},\ \bibinfo {pages} {094012} (\bibinfo {year} {2013})},\
  \Eprint {http://arxiv.org/abs/1309.0301} {arXiv:1309.0301 [hep-ph]}
  \BibitemShut {NoStop}%
\bibitem [{\citenamefont {Bernlochner}\ \emph {et~al.}(2017)\citenamefont
  {Bernlochner}, \citenamefont {Ligeti}, \citenamefont {Papucci},\ and\
  \citenamefont {Robinson}}]{Bernlochner:2017jka}%
  \BibitemOpen
  \bibfield  {author} {\bibinfo {author} {\bibfnamefont {F.~U.}\ \bibnamefont
  {Bernlochner}}, \bibinfo {author} {\bibfnamefont {Z.}~\bibnamefont {Ligeti}},
  \bibinfo {author} {\bibfnamefont {M.}~\bibnamefont {Papucci}}, \ and\
  \bibinfo {author} {\bibfnamefont {D.~J.}\ \bibnamefont {Robinson}},\
  }\bibfield  {title} {\enquote {\bibinfo {title} {\emph{Combined Analysis of
  Semileptonic $B$ decays to $D$ and $D^*$: $R(D^{(*)})$, $|V_{cb}|$, and New
  Physics}},}\ }\href {\doibase 10.1103/PhysRevD.95.115008} {\bibfield
  {journal} {\bibinfo  {journal} {Phys. Rev.}\ }\textbf {\bibinfo {volume}
  {D95}},\ \bibinfo {pages} {115008} (\bibinfo {year} {2017})},\ \Eprint
  {http://arxiv.org/abs/1703.05330} {arXiv:1703.05330 [hep-ph]} \BibitemShut
  {NoStop}%
\bibitem [{\citenamefont {Bigi}\ \emph {et~al.}(2017)\citenamefont {Bigi},
  \citenamefont {Gambino},\ and\ \citenamefont {Schacht}}]{Bigi:2017jbd}%
  \BibitemOpen
  \bibfield  {author} {\bibinfo {author} {\bibfnamefont {D.}~\bibnamefont
  {Bigi}}, \bibinfo {author} {\bibfnamefont {P.}~\bibnamefont {Gambino}}, \
  and\ \bibinfo {author} {\bibfnamefont {S.}~\bibnamefont {Schacht}},\
  }\bibfield  {title} {\enquote {\bibinfo {title} {\emph{$R(D^*)$, $|V_{cb}|$,
  and the Heavy Quark Symmetry Relations Between Form Factors}},}\ }\href
  {\doibase 10.1007/JHEP11(2017)061} {\bibfield  {journal} {\bibinfo  {journal}
  {JHEP}\ }\textbf {\bibinfo {volume} {11}},\ \bibinfo {pages} {061} (\bibinfo
  {year} {2017})},\ \Eprint {http://arxiv.org/abs/1707.09509} {arXiv:1707.09509
  [hep-ph]} \BibitemShut {NoStop}%
\bibitem [{\citenamefont {Jaiswal}\ \emph {et~al.}(2017)\citenamefont
  {Jaiswal}, \citenamefont {Nandi},\ and\ \citenamefont
  {Patra}}]{Jaiswal:2017rve}%
  \BibitemOpen
  \bibfield  {author} {\bibinfo {author} {\bibfnamefont {S.}~\bibnamefont
  {Jaiswal}}, \bibinfo {author} {\bibfnamefont {S.}~\bibnamefont {Nandi}}, \
  and\ \bibinfo {author} {\bibfnamefont {S.~K.}\ \bibnamefont {Patra}},\
  }\bibfield  {title} {\enquote {\bibinfo {title} {\emph{Extraction of
  $|V_{cb}|$ From $B\to D^{(*)}\ell\nu_\ell$ and the Standard Model Predictions
  of $R(D^{(*)})$}},}\ }\href@noop {} {\  (\bibinfo {year} {2017})},\ \Eprint
  {http://arxiv.org/abs/1707.09977} {arXiv:1707.09977 [hep-ph]} \BibitemShut
  {NoStop}%
\bibitem [{\citenamefont {Fajfer}\ \emph {et~al.}(2012)\citenamefont {Fajfer},
  \citenamefont {Kamenik},\ and\ \citenamefont {Nisandzic}}]{Fajfer:2012vx}%
  \BibitemOpen
  \bibfield  {author} {\bibinfo {author} {\bibfnamefont {S.}~\bibnamefont
  {Fajfer}}, \bibinfo {author} {\bibfnamefont {J.~F.}\ \bibnamefont {Kamenik}},
  \ and\ \bibinfo {author} {\bibfnamefont {I.}~\bibnamefont {Nisandzic}},\
  }\bibfield  {title} {\enquote {\bibinfo {title} {\emph{On the $B \to D^* \tau
  \bar \nu_{\tau}$ Sensitivity to New Physics}},}\ }\href {\doibase
  10.1103/PhysRevD.85.094025} {\bibfield  {journal} {\bibinfo  {journal} {Phys.
  Rev.}\ }\textbf {\bibinfo {volume} {D85}},\ \bibinfo {pages} {094025}
  (\bibinfo {year} {2012})},\ \Eprint {http://arxiv.org/abs/1203.2654}
  {arXiv:1203.2654 [hep-ph]} \BibitemShut {NoStop}%
\bibitem [{\citenamefont {Becirevic}\ \emph {et~al.}(2012)\citenamefont
  {Becirevic}, \citenamefont {Kosnik},\ and\ \citenamefont
  {Tayduganov}}]{Becirevic:2012jf}%
  \BibitemOpen
  \bibfield  {author} {\bibinfo {author} {\bibfnamefont {D.}~\bibnamefont
  {Becirevic}}, \bibinfo {author} {\bibfnamefont {N.}~\bibnamefont {Kosnik}}, \
  and\ \bibinfo {author} {\bibfnamefont {A.}~\bibnamefont {Tayduganov}},\
  }\bibfield  {title} {\enquote {\bibinfo {title} {\emph{$\bar B\to D\tau\bar
  \nu_\tau$ vs. $\bar B\to D\mu\bar \nu_\mu$}},}\ }\href {\doibase
  10.1016/j.physletb.2012.08.016} {\bibfield  {journal} {\bibinfo  {journal}
  {Phys. Lett.}\ }\textbf {\bibinfo {volume} {B716}},\ \bibinfo {pages}
  {208--213} (\bibinfo {year} {2012})},\ \Eprint
  {http://arxiv.org/abs/1206.4977} {arXiv:1206.4977 [hep-ph]} \BibitemShut
  {NoStop}%
\bibitem [{\citenamefont {Matyja}\ \emph {et~al.}(2007)\citenamefont {Matyja}
  \emph {et~al.}}]{Matyja:2007kt}%
  \BibitemOpen
  \bibfield  {author} {\bibinfo {author} {\bibfnamefont {A.}~\bibnamefont
  {Matyja}} \emph {et~al.} (\bibinfo {collaboration} {Belle}),\ }\bibfield
  {title} {\enquote {\bibinfo {title} {\emph{Observation of $B^0 \to D^{*-}
  \tau^+ \nu_\tau$ Decay at Belle}},}\ }\href {\doibase
  10.1103/PhysRevLett.99.191807} {\bibfield  {journal} {\bibinfo  {journal}
  {Phys. Rev. Lett.}\ }\textbf {\bibinfo {volume} {99}},\ \bibinfo {pages}
  {191807} (\bibinfo {year} {2007})},\ \Eprint {http://arxiv.org/abs/0706.4429}
  {arXiv:0706.4429 [hep-ex]} \BibitemShut {NoStop}%
\bibitem [{\citenamefont {Bozek}\ \emph {et~al.}(2010)\citenamefont {Bozek}
  \emph {et~al.}}]{Bozek:2010xy}%
  \BibitemOpen
  \bibfield  {author} {\bibinfo {author} {\bibfnamefont {A.}~\bibnamefont
  {Bozek}} \emph {et~al.} (\bibinfo {collaboration} {Belle}),\ }\bibfield
  {title} {\enquote {\bibinfo {title} {\emph{Observation of $B^+ \to
  \overline{D}^{*0} \tau^+ \nu_\tau$ and Evidence for $B^+\to \overline{D}^0
  \tau^+ \nu_\tau$ at Belle}},}\ }\href {\doibase 10.1103/PhysRevD.82.072005}
  {\bibfield  {journal} {\bibinfo  {journal} {Phys. Rev.}\ }\textbf {\bibinfo
  {volume} {D82}},\ \bibinfo {pages} {072005} (\bibinfo {year} {2010})},\
  \Eprint {http://arxiv.org/abs/1005.2302} {arXiv:1005.2302 [hep-ex]}
  \BibitemShut {NoStop}%
\bibitem [{\citenamefont {Hirose}\ \emph {et~al.}(2017)\citenamefont {Hirose}
  \emph {et~al.}}]{Hirose:2016wfn}%
  \BibitemOpen
  \bibfield  {author} {\bibinfo {author} {\bibfnamefont {S.}~\bibnamefont
  {Hirose}} \emph {et~al.} (\bibinfo {collaboration} {Belle}),\ }\bibfield
  {title} {\enquote {\bibinfo {title} {\emph{Measurement of the $\tau$ Lepton
  Polarization and $R(D^*)$ in the Decay $\bar{B} \to D^* \tau^-
  \bar{\nu}_\tau$}},}\ }\href {\doibase 10.1103/PhysRevLett.118.211801}
  {\bibfield  {journal} {\bibinfo  {journal} {Phys. Rev. Lett.}\ }\textbf
  {\bibinfo {volume} {118}},\ \bibinfo {pages} {211801} (\bibinfo {year}
  {2017})},\ \Eprint {http://arxiv.org/abs/1612.00529} {arXiv:1612.00529
  [hep-ex]} \BibitemShut {NoStop}%
\bibitem [{\citenamefont {Lees}\ \emph {et~al.}(2012)\citenamefont {Lees} \emph
  {et~al.}}]{Lees:2012xj}%
  \BibitemOpen
  \bibfield  {author} {\bibinfo {author} {\bibfnamefont {J.~P.}\ \bibnamefont
  {Lees}} \emph {et~al.} (\bibinfo {collaboration} {BaBar}),\ }\bibfield
  {title} {\enquote {\bibinfo {title} {\emph{Evidence for an Excess of $\bar{B}
  \to D^{(*)} \tau^-\bar{\nu}_\tau$ Decays}},}\ }\href {\doibase
  10.1103/PhysRevLett.109.101802} {\bibfield  {journal} {\bibinfo  {journal}
  {Phys. Rev. Lett.}\ }\textbf {\bibinfo {volume} {109}},\ \bibinfo {pages}
  {101802} (\bibinfo {year} {2012})},\ \Eprint {http://arxiv.org/abs/1205.5442}
  {arXiv:1205.5442 [hep-ex]} \BibitemShut {NoStop}%
\bibitem [{\citenamefont {Lees}\ \emph {et~al.}(2013)\citenamefont {Lees} \emph
  {et~al.}}]{Lees:2013uzd}%
  \BibitemOpen
  \bibfield  {author} {\bibinfo {author} {\bibfnamefont {J.~P.}\ \bibnamefont
  {Lees}} \emph {et~al.} (\bibinfo {collaboration} {BaBar}),\ }\bibfield
  {title} {\enquote {\bibinfo {title} {\emph{Measurement of an Excess of
  $\bar{B} \to D^{(*)}\tau^- \bar{\nu}_\tau$ Decays and Implications for
  Charged Higgs Bosons}},}\ }\href {\doibase 10.1103/PhysRevD.88.072012}
  {\bibfield  {journal} {\bibinfo  {journal} {Phys. Rev.}\ }\textbf {\bibinfo
  {volume} {D88}},\ \bibinfo {pages} {072012} (\bibinfo {year} {2013})},\
  \Eprint {http://arxiv.org/abs/1303.0571} {arXiv:1303.0571 [hep-ex]}
  \BibitemShut {NoStop}%
\bibitem [{\citenamefont {Aaij}\ \emph {et~al.}(2015)\citenamefont {Aaij} \emph
  {et~al.}}]{Aaij:2015yra}%
  \BibitemOpen
  \bibfield  {author} {\bibinfo {author} {\bibfnamefont {R.}~\bibnamefont
  {Aaij}} \emph {et~al.} (\bibinfo {collaboration} {LHCb}),\ }\bibfield
  {title} {\enquote {\bibinfo {title} {\emph{Measurement of the Ratio of
  Branching Fractions $\mathcal{B}(\bar{B}^0 \to
  D^{*+}\tau^{-}\bar{\nu}_{\tau})/\mathcal{B}(\bar{B}^0 \to
  D^{*+}\mu^{-}\bar{\nu}_{\mu})$}},}\ }\href {\doibase
  10.1103/PhysRevLett.115.159901, 10.1103/PhysRevLett.115.111803} {\bibfield
  {journal} {\bibinfo  {journal} {Phys. Rev. Lett.}\ }\textbf {\bibinfo
  {volume} {115}},\ \bibinfo {pages} {111803} (\bibinfo {year} {2015})},\
  \bibinfo {note} {[Erratum: Phys. Rev. Lett.115, no.15, 159901 (2015)]},\
  \Eprint {http://arxiv.org/abs/1506.08614} {arXiv:1506.08614 [hep-ex]}
  \BibitemShut {NoStop}%
\bibitem [{\citenamefont {Amhis}\ \emph {et~al.}(2016)\citenamefont {Amhis}
  \emph {et~al.}}]{Amhis:2016xyh}%
  \BibitemOpen
  \bibfield  {author} {\bibinfo {author} {\bibfnamefont {Y.}~\bibnamefont
  {Amhis}} \emph {et~al.},\ }\bibfield  {title} {\enquote {\bibinfo {title}
  {\emph{Averages of $b$-Hadron, $c$-Hadron, and $\tau$-Lepton Properties as of
  Summer 2016}},}\ }\href@noop {} {\  (\bibinfo {year} {2016})},\ \Eprint
  {http://arxiv.org/abs/1612.07233} {arXiv:1612.07233 [hep-ex]} \BibitemShut
  {NoStop}%
\bibitem [{\citenamefont {Goldberger}(1999)}]{Goldberger:1999yh}%
  \BibitemOpen
  \bibfield  {author} {\bibinfo {author} {\bibfnamefont {W.~D.}\ \bibnamefont
  {Goldberger}},\ }\bibfield  {title} {\enquote {\bibinfo {title}
  {\emph{Semileptonic $B$ Decays as a Probe of New Physics}},}\ }\href@noop {}
  {\  (\bibinfo {year} {1999})},\ \Eprint {http://arxiv.org/abs/hep-ph/9902311}
  {arXiv:hep-ph/9902311 [hep-ph]} \BibitemShut {NoStop}%
\bibitem [{\citenamefont {Cirigliano}\ \emph {et~al.}(2013)\citenamefont
  {Cirigliano}, \citenamefont {Gonzalez-Alonso},\ and\ \citenamefont
  {Graesser}}]{Cirigliano:2012ab}%
  \BibitemOpen
  \bibfield  {author} {\bibinfo {author} {\bibfnamefont {V.}~\bibnamefont
  {Cirigliano}}, \bibinfo {author} {\bibfnamefont {M.}~\bibnamefont
  {Gonzalez-Alonso}}, \ and\ \bibinfo {author} {\bibfnamefont {M.~L.}\
  \bibnamefont {Graesser}},\ }\bibfield  {title} {\enquote {\bibinfo {title}
  {\emph{Non-Standard Charged Current Interactions: Beta Decays Versus the
  LHC}},}\ }\href {\doibase 10.1007/JHEP02(2013)046} {\bibfield  {journal}
  {\bibinfo  {journal} {JHEP}\ }\textbf {\bibinfo {volume} {02}},\ \bibinfo
  {pages} {046} (\bibinfo {year} {2013})},\ \Eprint
  {http://arxiv.org/abs/1210.4553} {arXiv:1210.4553 [hep-ph]} \BibitemShut
  {NoStop}%
\bibitem [{\citenamefont {Alonso}\ \emph {et~al.}(2017)\citenamefont {Alonso},
  \citenamefont {Grinstein},\ and\ \citenamefont
  {Martin~Camalich}}]{Alonso:2016oyd}%
  \BibitemOpen
  \bibfield  {author} {\bibinfo {author} {\bibfnamefont {R.}~\bibnamefont
  {Alonso}}, \bibinfo {author} {\bibfnamefont {B.}~\bibnamefont {Grinstein}}, \
  and\ \bibinfo {author} {\bibfnamefont {J.}~\bibnamefont {Martin~Camalich}},\
  }\bibfield  {title} {\enquote {\bibinfo {title} {\emph{Lifetime of $B_c^-$
  Constrains Explanations for Anomalies in $B\to D^{(*)}\tau\nu$}},}\ }\href
  {\doibase 10.1103/PhysRevLett.118.081802} {\bibfield  {journal} {\bibinfo
  {journal} {Phys. Rev. Lett.}\ }\textbf {\bibinfo {volume} {118}},\ \bibinfo
  {pages} {081802} (\bibinfo {year} {2017})},\ \Eprint
  {http://arxiv.org/abs/1611.06676} {arXiv:1611.06676 [hep-ph]} \BibitemShut
  {NoStop}%
\bibitem [{\citenamefont {Akeroyd}\ and\ \citenamefont
  {Chen}(2017)}]{Akeroyd:2017mhr}%
  \BibitemOpen
  \bibfield  {author} {\bibinfo {author} {\bibfnamefont {A.~G.}\ \bibnamefont
  {Akeroyd}}\ and\ \bibinfo {author} {\bibfnamefont {C.-H.}\ \bibnamefont
  {Chen}},\ }\bibfield  {title} {\enquote {\bibinfo {title} {\emph{Constraint
  on the Branching Ratio of $B_c \to \tau \nu$ From LEP1 and Consequences for
  $R_{D^{(*)}}$ Anomaly}},}\ }\href {\doibase 10.1103/PhysRevD.96.075011}
  {\bibfield  {journal} {\bibinfo  {journal} {Phys. Rev.}\ }\textbf {\bibinfo
  {volume} {D96}},\ \bibinfo {pages} {075011} (\bibinfo {year} {2017})},\
  \Eprint {http://arxiv.org/abs/1708.04072} {arXiv:1708.04072 [hep-ph]}
  \BibitemShut {NoStop}%
\bibitem [{\citenamefont {Valencia}\ and\ \citenamefont
  {Willenbrock}(1994)}]{Valencia:1994cj}%
  \BibitemOpen
  \bibfield  {author} {\bibinfo {author} {\bibfnamefont {G.}~\bibnamefont
  {Valencia}}\ and\ \bibinfo {author} {\bibfnamefont {S.}~\bibnamefont
  {Willenbrock}},\ }\bibfield  {title} {\enquote {\bibinfo {title} {\emph{Quark
  - Lepton Unification and Rare Meson Decays}},}\ }\href {\doibase
  10.1103/PhysRevD.50.6843} {\bibfield  {journal} {\bibinfo  {journal} {Phys.
  Rev.}\ }\textbf {\bibinfo {volume} {D50}},\ \bibinfo {pages} {6843--6848}
  (\bibinfo {year} {1994})},\ \Eprint {http://arxiv.org/abs/hep-ph/9409201}
  {arXiv:hep-ph/9409201 [hep-ph]} \BibitemShut {NoStop}%
\bibitem [{\citenamefont {Smirnov}(2007)}]{Smirnov:2007hv}%
  \BibitemOpen
  \bibfield  {author} {\bibinfo {author} {\bibfnamefont {A.~D.}\ \bibnamefont
  {Smirnov}},\ }\bibfield  {title} {\enquote {\bibinfo {title} {\emph{Mass
  Limits for Scalar and Gauge Leptoquarks from $K_L^0 \rightarrow e^\mp
  \mu^\pm$, $B^0 \rightarrow e^\mp \tau^\pm$ Decays}},}\ }\href {\doibase
  10.1142/S0217732307024401} {\bibfield  {journal} {\bibinfo  {journal} {Mod.
  Phys. Lett.}\ }\textbf {\bibinfo {volume} {A22}},\ \bibinfo {pages}
  {2353--2363} (\bibinfo {year} {2007})},\ \Eprint
  {http://arxiv.org/abs/0705.0308} {arXiv:0705.0308 [hep-ph]} \BibitemShut
  {NoStop}%
\bibitem [{\citenamefont {Smirnov}(2008)}]{Smirnov:2008zzb}%
  \BibitemOpen
  \bibfield  {author} {\bibinfo {author} {\bibfnamefont {A.~D.}\ \bibnamefont
  {Smirnov}},\ }\bibfield  {title} {\enquote {\bibinfo {title}
  {\emph{Contributions of Gauge and Scalar Leptoquarks to $K_L^0 \rightarrow
  l_i^+ l_j^-$ and $B^0 \rightarrow l_i^+ l_j^-$ Decay and Constraints on
  Leptoquark Masses from the Decays $K_L^0 \rightarrow e^\mp \mu^\pm$ and $B^0
  \rightarrow e^\mp \tau^\pm$}},}\ }\href {\doibase 10.1134/S106377880808019X}
  {\bibfield  {journal} {\bibinfo  {journal} {Phys. Atom. Nucl.}\ }\textbf
  {\bibinfo {volume} {71}},\ \bibinfo {pages} {1470--1480} (\bibinfo {year}
  {2008})},\ \bibinfo {note} {[Yad. Fiz.71,1498(2008)]}\BibitemShut {NoStop}%
\bibitem [{\citenamefont {Carpentier}\ and\ \citenamefont
  {Davidson}(2010)}]{Carpentier:2010ue}%
  \BibitemOpen
  \bibfield  {author} {\bibinfo {author} {\bibfnamefont {M.}~\bibnamefont
  {Carpentier}}\ and\ \bibinfo {author} {\bibfnamefont {S.}~\bibnamefont
  {Davidson}},\ }\bibfield  {title} {\enquote {\bibinfo {title}
  {\emph{Constraints on Two-Lepton, Two Quark Operators}},}\ }\href {\doibase
  10.1140/epjc/s10052-010-1482-4} {\bibfield  {journal} {\bibinfo  {journal}
  {Eur. Phys. J.}\ }\textbf {\bibinfo {volume} {C70}},\ \bibinfo {pages}
  {1071--1090} (\bibinfo {year} {2010})},\ \Eprint
  {http://arxiv.org/abs/1008.0280} {arXiv:1008.0280 [hep-ph]} \BibitemShut
  {NoStop}%
\bibitem [{\citenamefont {Kuznetsov}\ \emph {et~al.}(2012)\citenamefont
  {Kuznetsov}, \citenamefont {Mikheev},\ and\ \citenamefont
  {Serghienko}}]{Kuznetsov:2012ai}%
  \BibitemOpen
  \bibfield  {author} {\bibinfo {author} {\bibfnamefont {A.~V.}\ \bibnamefont
  {Kuznetsov}}, \bibinfo {author} {\bibfnamefont {N.~V.}\ \bibnamefont
  {Mikheev}}, \ and\ \bibinfo {author} {\bibfnamefont {A.~V.}\ \bibnamefont
  {Serghienko}},\ }\bibfield  {title} {\enquote {\bibinfo {title} {\emph{The
  Third Type of Fermion Mixing in the Lepton and Quark Interactions with
  Leptoquarks}},}\ }\href {\doibase 10.1142/S0217751X12500625} {\bibfield
  {journal} {\bibinfo  {journal} {Int. J. Mod. Phys.}\ }\textbf {\bibinfo
  {volume} {A27}},\ \bibinfo {pages} {1250062} (\bibinfo {year} {2012})},\
  \Eprint {http://arxiv.org/abs/1203.0196} {arXiv:1203.0196 [hep-ph]}
  \BibitemShut {NoStop}%
\bibitem [{\citenamefont {Aubert}\ \emph {et~al.}(2006)\citenamefont {Aubert}
  \emph {et~al.}}]{Aubert:2006vb}%
  \BibitemOpen
  \bibfield  {author} {\bibinfo {author} {\bibfnamefont {B.}~\bibnamefont
  {Aubert}} \emph {et~al.} (\bibinfo {collaboration} {BaBar}),\ }\bibfield
  {title} {\enquote {\bibinfo {title} {\emph{Measurements of Branching
  Fractions, Rate Asymmetries, and Angular Distributions in the Rare Decays $B
  \to K \ell^{+} \ell^{-}$ and $B \to K^{*} \ell^{+} \ell^{-}$}},}\ }\href
  {\doibase 10.1103/PhysRevD.73.092001} {\bibfield  {journal} {\bibinfo
  {journal} {Phys. Rev.}\ }\textbf {\bibinfo {volume} {D73}},\ \bibinfo {pages}
  {092001} (\bibinfo {year} {2006})},\ \Eprint
  {http://arxiv.org/abs/hep-ex/0604007} {arXiv:hep-ex/0604007 [hep-ex]}
  \BibitemShut {NoStop}%
\bibitem [{\citenamefont {Baldini}\ \emph {et~al.}(2016)\citenamefont {Baldini}
  \emph {et~al.}}]{TheMEG:2016wtm}%
  \BibitemOpen
  \bibfield  {author} {\bibinfo {author} {\bibfnamefont {A.~M.}\ \bibnamefont
  {Baldini}} \emph {et~al.} (\bibinfo {collaboration} {MEG}),\ }\bibfield
  {title} {\enquote {\bibinfo {title} {\emph{Search for the Lepton Flavour
  Violating Decay $\mu ^+ \rightarrow {e}^+ \gamma $ with the Full Dataset of
  the MEG Experiment}},}\ }\href {\doibase 10.1140/epjc/s10052-016-4271-x}
  {\bibfield  {journal} {\bibinfo  {journal} {Eur. Phys. J.}\ }\textbf
  {\bibinfo {volume} {C76}},\ \bibinfo {pages} {434} (\bibinfo {year}
  {2016})},\ \Eprint {http://arxiv.org/abs/1605.05081} {arXiv:1605.05081
  [hep-ex]} \BibitemShut {NoStop}%
\bibitem [{\citenamefont {Fornal}\ and\ \citenamefont {Grinstein}()}]{work}%
  \BibitemOpen
  \bibfield  {author} {\bibinfo {author} {\bibfnamefont {B.}~\bibnamefont
  {Fornal}}\ and\ \bibinfo {author} {\bibfnamefont {B.}~\bibnamefont
  {Grinstein}},\ }\href@noop {} {\bibinfo  {journal} {work in progress}\
  }\BibitemShut {NoStop}%
\bibitem [{\citenamefont {Calibbi}\ \emph {et~al.}(2017)\citenamefont
  {Calibbi}, \citenamefont {Crivellin},\ and\ \citenamefont
  {Li}}]{Calibbi:2017qbu}%
  \BibitemOpen
\bibfield  {journal} {  }\bibfield  {author} {\bibinfo {author} {\bibfnamefont
  {L.}~\bibnamefont {Calibbi}}, \bibinfo {author} {\bibfnamefont
  {A.}~\bibnamefont {Crivellin}}, \ and\ \bibinfo {author} {\bibfnamefont
  {T.}~\bibnamefont {Li}},\ }\bibfield  {title} {\enquote {\bibinfo {title}
  {\emph{A Model of Vector Leptoquarks in View of the $B$-Physics
  Anomalies}},}\ }\href@noop {} {\  (\bibinfo {year} {2017})},\ \Eprint
  {http://arxiv.org/abs/1709.00692} {arXiv:1709.00692 [hep-ph]} \BibitemShut
  {NoStop}%
\bibitem [{\citenamefont {Di~Luzio}\ \emph {et~al.}(2017)\citenamefont
  {Di~Luzio}, \citenamefont {Greljo},\ and\ \citenamefont
  {Nardecchia}}]{DiLuzio:2017vat}%
  \BibitemOpen
  \bibfield  {author} {\bibinfo {author} {\bibfnamefont {L.}~\bibnamefont
  {Di~Luzio}}, \bibinfo {author} {\bibfnamefont {A.}~\bibnamefont {Greljo}}, \
  and\ \bibinfo {author} {\bibfnamefont {M.}~\bibnamefont {Nardecchia}},\
  }\bibfield  {title} {\enquote {\bibinfo {title} {\emph{Gauge Leptoquark as
  the Origin of $B$-Physics Anomalies}},}\ }\href@noop {} {\  (\bibinfo {year}
  {2017})},\ \Eprint {http://arxiv.org/abs/1708.08450} {arXiv:1708.08450
  [hep-ph]} \BibitemShut {NoStop}%
\bibitem [{\citenamefont {Slansky}(1981)}]{Slansky:1981yr}%
  \BibitemOpen
  \bibfield  {author} {\bibinfo {author} {\bibfnamefont {R.}~\bibnamefont
  {Slansky}},\ }\bibfield  {title} {\enquote {\bibinfo {title} {\emph{Group
  Theory for Unified Model Building}},}\ }\href {\doibase
  10.1016/0370-1573(81)90092-2} {\bibfield  {journal} {\bibinfo  {journal}
  {Phys. Rept.}\ }\textbf {\bibinfo {volume} {79}},\ \bibinfo {pages} {1--128}
  (\bibinfo {year} {1981})}\BibitemShut {NoStop}%
\bibitem [{\citenamefont {Sirunyan}\ \emph {et~al.}(2017)\citenamefont
  {Sirunyan} \emph {et~al.}}]{Sirunyan:2016iap}%
  \BibitemOpen
  \bibfield  {author} {\bibinfo {author} {\bibfnamefont {A.~M}\ \bibnamefont
  {Sirunyan}} \emph {et~al.} (\bibinfo {collaboration} {CMS}),\ }\bibfield
  {title} {\enquote {\bibinfo {title} {\emph{Search for Dijet Resonances in
  Proton-Proton Collisions at $\sqrt{s}$ = 13 TeV and Constraints on Dark
  Matter and Other Models}},}\ }\href {\doibase 10.1016/j.physletb.2017.02.012}
  {\bibfield  {journal} {\bibinfo  {journal} {Phys. Lett.}\ }\textbf {\bibinfo
  {volume} {B769}},\ \bibinfo {pages} {520--542} (\bibinfo {year} {2017})},\
  \Eprint {http://arxiv.org/abs/1611.03568} {arXiv:1611.03568 [hep-ex]}
  \BibitemShut {NoStop}%
\bibitem [{\citenamefont {Aaboud}\ \emph {et~al.}(2017)\citenamefont {Aaboud}
  \emph {et~al.}}]{Aaboud:2017yvp}%
  \BibitemOpen
  \bibfield  {author} {\bibinfo {author} {\bibfnamefont {M.}~\bibnamefont
  {Aaboud}} \emph {et~al.} (\bibinfo {collaboration} {ATLAS}),\ }\bibfield
  {title} {\enquote {\bibinfo {title} {\emph{Search for New Phenomena in Dijet
  Events Using 37 fb$^{-1}$ of $pp$ Collision Data Collected at $\sqrt{s}=$13
  TeV with the ATLAS Detector}},}\ }\href {\doibase 10.1103/PhysRevD.96.052004}
  {\bibfield  {journal} {\bibinfo  {journal} {Phys. Rev.}\ }\textbf {\bibinfo
  {volume} {D96}},\ \bibinfo {pages} {052004} (\bibinfo {year} {2017})},\
  \Eprint {http://arxiv.org/abs/1703.09127} {arXiv:1703.09127 [hep-ex]}
  \BibitemShut {NoStop}%
\bibitem [{\citenamefont {Isidori}\ \emph {et~al.}(2010)\citenamefont
  {Isidori}, \citenamefont {Nir},\ and\ \citenamefont
  {Perez}}]{Isidori:2010kg}%
  \BibitemOpen
  \bibfield  {author} {\bibinfo {author} {\bibfnamefont {G.}~\bibnamefont
  {Isidori}}, \bibinfo {author} {\bibfnamefont {Y.}~\bibnamefont {Nir}}, \ and\
  \bibinfo {author} {\bibfnamefont {G.}~\bibnamefont {Perez}},\ }\bibfield
  {title} {\enquote {\bibinfo {title} {\emph{Flavor Physics Constraints for
  Physics Beyond the Standard Model}},}\ }\href {\doibase
  10.1146/annurev.nucl.012809.104534} {\bibfield  {journal} {\bibinfo
  {journal} {Ann. Rev. Nucl. Part. Sci.}\ }\textbf {\bibinfo {volume} {60}},\
  \bibinfo {pages} {355} (\bibinfo {year} {2010})},\ \Eprint
  {http://arxiv.org/abs/1002.0900} {arXiv:1002.0900 [hep-ph]} \BibitemShut
  {NoStop}%
\bibitem [{\citenamefont {Blake}\ \emph {et~al.}(2017)\citenamefont {Blake},
  \citenamefont {Lanfranchi},\ and\ \citenamefont {Straub}}]{Blake:2016olu}%
  \BibitemOpen
  \bibfield  {author} {\bibinfo {author} {\bibfnamefont {T.}~\bibnamefont
  {Blake}}, \bibinfo {author} {\bibfnamefont {G.}~\bibnamefont {Lanfranchi}}, \
  and\ \bibinfo {author} {\bibfnamefont {D.~M.}\ \bibnamefont {Straub}},\
  }\bibfield  {title} {\enquote {\bibinfo {title} {\emph{Rare $B$ Decays as
  Tests of the Standard Model}},}\ }\href {\doibase 10.1016/j.ppnp.2016.10.001}
  {\bibfield  {journal} {\bibinfo  {journal} {Prog. Part. Nucl. Phys.}\
  }\textbf {\bibinfo {volume} {92}},\ \bibinfo {pages} {50--91} (\bibinfo
  {year} {2017})},\ \Eprint {http://arxiv.org/abs/1606.00916} {arXiv:1606.00916
  [hep-ph]} \BibitemShut {NoStop}%
\bibitem [{\citenamefont {Maalampi}\ \emph {et~al.}(1988)\citenamefont
  {Maalampi}, \citenamefont {Pietila},\ and\ \citenamefont
  {Vilja}}]{Maalampi:1987pb}%
  \BibitemOpen
  \bibfield  {author} {\bibinfo {author} {\bibfnamefont {J.}~\bibnamefont
  {Maalampi}}, \bibinfo {author} {\bibfnamefont {A.}~\bibnamefont {Pietila}}, \
  and\ \bibinfo {author} {\bibfnamefont {I.}~\bibnamefont {Vilja}},\ }\bibfield
   {title} {\enquote {\bibinfo {title} {\emph{Constraints for the Diquark
  Couplings from $q\rightarrow q + \gamma$}},}\ }\href {\doibase
  10.1142/S0217732388000465} {\bibfield  {journal} {\bibinfo  {journal} {Mod.
  Phys. Lett.}\ }\textbf {\bibinfo {volume} {3A}},\ \bibinfo {pages} {373}
  (\bibinfo {year} {1988})}\BibitemShut {NoStop}%
\bibitem [{\citenamefont {Chakraverty}\ and\ \citenamefont
  {Choudhury}(2001)}]{Chakraverty:2000rm}%
  \BibitemOpen
  \bibfield  {author} {\bibinfo {author} {\bibfnamefont {D.}~\bibnamefont
  {Chakraverty}}\ and\ \bibinfo {author} {\bibfnamefont {D.}~\bibnamefont
  {Choudhury}},\ }\bibfield  {title} {\enquote {\bibinfo {title} {\emph{$b \to
  s \gamma$ Confronts $B$ Violating Scalar Couplings: R-parity Violating
  Supersymmetry or Diquark}},}\ }\href {\doibase 10.1103/PhysRevD.63.075009}
  {\bibfield  {journal} {\bibinfo  {journal} {Phys. Rev.}\ }\textbf {\bibinfo
  {volume} {D63}},\ \bibinfo {pages} {075009} (\bibinfo {year} {2001})},\
  \Eprint {http://arxiv.org/abs/hep-ph/0008165} {arXiv:hep-ph/0008165 [hep-ph]}
  \BibitemShut {NoStop}%
\bibitem [{\citenamefont {Phillips}\ \emph {et~al.}(2016)\citenamefont
  {Phillips} \emph {et~al.}}]{Phillips:2014fgb}%
  \BibitemOpen
  \bibfield  {author} {\bibinfo {author} {\bibfnamefont {D.~G.}\ \bibnamefont
  {Phillips}, \bibfnamefont {II}} \emph {et~al.},\ }\bibfield  {title}
  {\enquote {\bibinfo {title} {\emph{Neutron-Antineutron Oscillations:
  Theoretical Status and Experimental Prospects}},}\ }\href {\doibase
  10.1016/j.physrep.2015.11.001} {\bibfield  {journal} {\bibinfo  {journal}
  {Phys. Rept.}\ }\textbf {\bibinfo {volume} {612}},\ \bibinfo {pages} {1--45}
  (\bibinfo {year} {2016})},\ \Eprint {http://arxiv.org/abs/1410.1100}
  {arXiv:1410.1100 [hep-ex]} \BibitemShut {NoStop}%
\bibitem [{\citenamefont {Minkowski}(1977)}]{Minkowski:1977sc}%
  \BibitemOpen
  \bibfield  {author} {\bibinfo {author} {\bibfnamefont {P.}~\bibnamefont
  {Minkowski}},\ }\bibfield  {title} {\enquote {\bibinfo {title} {\emph{$\mu
  \to e\gamma$ at a Rate of One Out of $10^{9}$ Muon Decays?}}}\ }\href
  {\doibase 10.1016/0370-2693(77)90435-X} {\bibfield  {journal} {\bibinfo
  {journal} {Phys. Lett.}\ }\textbf {\bibinfo {volume} {67B}},\ \bibinfo
  {pages} {421--428} (\bibinfo {year} {1977})}\BibitemShut {NoStop}%
\bibitem [{\citenamefont {Tsutsui}\ \emph {et~al.}(2004)\citenamefont {Tsutsui}
  \emph {et~al.}}]{Tsutsui:2004qc}%
  \BibitemOpen
  \bibfield  {author} {\bibinfo {author} {\bibfnamefont {N.}~\bibnamefont
  {Tsutsui}} \emph {et~al.} (\bibinfo {collaboration} {JLQCD, CP-PACS}),\
  }\bibfield  {title} {\enquote {\bibinfo {title} {\emph{Lattice QCD
  Calculation of the Proton Decay Matrix Element in the Continuum Limit}},}\
  }\href {\doibase 10.1103/PhysRevD.70.111501} {\bibfield  {journal} {\bibinfo
  {journal} {Phys. Rev.}\ }\textbf {\bibinfo {volume} {D70}},\ \bibinfo {pages}
  {111501} (\bibinfo {year} {2004})},\ \Eprint
  {http://arxiv.org/abs/hep-lat/0402026} {arXiv:hep-lat/0402026 [hep-lat]}
  \BibitemShut {NoStop}%
\bibitem [{\citenamefont {Abe}\ \emph {et~al.}(2015)\citenamefont {Abe} \emph
  {et~al.}}]{Abe:2011ky}%
  \BibitemOpen
  \bibfield  {author} {\bibinfo {author} {\bibfnamefont {K.}~\bibnamefont
  {Abe}} \emph {et~al.} (\bibinfo {collaboration} {Super-Kamiokande}),\
  }\bibfield  {title} {\enquote {\bibinfo {title} {\emph{The Search for
  $n-\bar{n}$ Oscillation in Super-Kamiokande I}},}\ }\href {\doibase
  10.1103/PhysRevD.91.072006} {\bibfield  {journal} {\bibinfo  {journal} {Phys.
  Rev.}\ }\textbf {\bibinfo {volume} {D91}},\ \bibinfo {pages} {072006}
  (\bibinfo {year} {2015})},\ \Eprint {http://arxiv.org/abs/1109.4227}
  {arXiv:1109.4227 [hep-ex]} \BibitemShut {NoStop}%
\bibitem [{\citenamefont {Stone}\ and\ \citenamefont
  {Uttayarat}(2012)}]{Stone:2011dn}%
  \BibitemOpen
  \bibfield  {author} {\bibinfo {author} {\bibfnamefont {D.~C.}\ \bibnamefont
  {Stone}}\ and\ \bibinfo {author} {\bibfnamefont {P.}~\bibnamefont
  {Uttayarat}},\ }\bibfield  {title} {\enquote {\bibinfo {title}
  {\emph{Explaining the $t\bar{t}$ Forward-Backward Asymmetry from a
  GUT-Inspired Model}},}\ }\href {\doibase 10.1007/JHEP01(2012)096} {\bibfield
  {journal} {\bibinfo  {journal} {JHEP}\ }\textbf {\bibinfo {volume} {01}},\
  \bibinfo {pages} {096} (\bibinfo {year} {2012})},\ \Eprint
  {http://arxiv.org/abs/1111.2050} {arXiv:1111.2050 [hep-ph]} \BibitemShut
  {NoStop}%
\end{thebibliography}%

\end{document}